\documentclass[fleqn,usenatbib]{mnras}

\usepackage{mathptmx}

\usepackage[T1]{fontenc}
\usepackage{ae,aecompl}

\usepackage{graphicx}	
\usepackage{amsmath}	
\usepackage{amssymb}	
\usepackage{bm}

\newcommand{\vektor}[1]{\bm{#1}}
\newcommand{\subrhoB}{{\rho\vektor{B}}}
\newcommand{\subvelB}{{\vektor{vB}}}
\newcommand{\vnabla}{\bm{\nabla}}

\newcommand{\percc}{\mathrm{cm}^{-3}}
\newcommand{\gpercc}{\mathrm{g\,cm}^{-3}}
\newcommand{\perscm}{\mathrm{cm}^{-2}}
\newcommand{\gperscm}{\mathrm{g\,cm}^{-2}}

\title[Magnetic field alignment in the ISM]{Alignment of the magnetic field in star forming regions and why it might be difficult to observe}

\author[Girichidis]{
Philipp Girichidis,$^{1,2}$\thanks{E-mail: philipp@girichidis.com}\\
$^{1}$Leibniz-Institut f\"{u}r Astrophysik Potsdam (AIP), An der Sternwarte 16, D-14482 Potsdam, Germany\\
$^{2}$Universit\"{a}t Heidelberg, Zentrum f\"{u}r Astronomie, Institut f\"{u}r Theoretische Astrophysik, Albert-Ueberle-Str. 2, D-69120 Heidelberg, Germany
}

\date{Accepted XXX. Received YYY; in original form ZZZ}

\pubyear{2020}

\begin{document}
\label{firstpage}
\pagerange{\pageref{firstpage}--\pageref{lastpage}}
\maketitle

\begin{abstract}
Magnetic fields are an important component of the interstellar medium (ISM) and exhibit strongly varying field strengths and a non-trivial correlation with the gas density. Its dynamical impact varies between individual regions of the ISM and correlates with the orientation of the field with respect to the gas structures. Using high-resolution magneto-hydrodynamical simulations of the ISM we explore the connection between the orientation of the field and the dynamical state of the gas. We find that the onset of gravitational instability in molecular gas above a density of $\rho\sim10^{-21}\,\gpercc$ $(n\sim400\,\percc)$ coincides with an alignment of the magnetic field lines and the gas flow. At this transition the gradient of the density changes from mainly perpendicular to preferentially parallel to the field lines. A connection between the three-dimensional alignment and projected two-dimensional observables is non-trivial, because of a large dispersion of the magnetic field orientation along the line of sight. The turbulent correlation lengths can be small compared to the typical integration lengths. As a consequence the small scale signal of the orientation can sensitively depend on the line of sight or the dynamical state of the cloud, can fluctuate stochastically or be completely averaged out. With higher spatial resolution more small scale structures are resolved, which aggravates the link between magneto-hydrodynamical quantities and projected observables.
\end{abstract}




\section{Introduction}

Magnetic fields are ubiquitously observed in nearby galaxies \citep{Beck2009, FletcherEtAl2011, Beck2012, Beck2015} as well as in the Milky Way \citep{Haverkorn2015}.  The field permeates the interstellar medium (ISM) with average field energy densities comparable to the thermal and cosmic ray component \citep{Cox2005}, which indicates that they can be dynamically relevant. Their importance for the formation of stars has been discussed controversially \citep{MestelSpitzer1956, Parker1966, MouschoviasSpitzer1976, Shu87, Elmegreen1989, HennebelleFalgarone2012, MacLowKlessen2004, McKeeOstriker2007, KlessenGlover2016}. The impact of magnetic fields on the structures of the galaxy and the ISM has been investigated with three-dimensional magneto-hydrodynamical simulations \citep{PadoanNordlund1999, deAvillezBreitschwerdt2005, Ziegler05, HennebelleEtAl2008, BanerjeeEtAl2009, FederrathEtAl2011, HillEtAl2012, PakmorSpringel2013, SteinwandelEtAl2019, KoertgenEtAl2019}. Recent reviews describe the theoretical efforts on galactic scales \citep{NaabOstriker2017}, in molecular clouds \citep{HennebelleInutsuka2019}, and in star forming regions \citep{KrumholzFederrath2019,WursterLi2019}.

In the warm atomic gas the magnetic energy might dominate over the gravitational energy \citep{Beck2001,HeilesTroland2005,Beck2012}. Ions couple to the magnetic field via the Lorentz force, whereas neutrals do not interact with the field. If the gas is sufficiently ionizied such that the collisions of ions and neutrals effectively transfer the effects of the Lorentz force also to the neutral component, the field lines are frozen in the gas. Combined with an infinite conductivity, the combined magneto-hydrodynamic (MHD) equations reduce to the ideal MHD approximation \citep[see, e.g.][]{GirichidisEtAl2020b}. For most of the volume in the ISM including large fractions of molecular clouds ideal MHD is a valid approximation. As a result, the magnetic pressure can support the gas against gravitational contraction \citep[e.g.][]{NakanoHishiUmebayashi2002,NtormousiEtAl2016}. Based on the mass-to-magnetic-flux ratio \citet{MestelSpitzer1956} concluded that the subcritical (i.e. magnetically supported) atomic clouds cannot collapse and form stars. Numerical simulations and observations reveal the opposite. However, the details of gravitational collapse in the presence of magnetic fields is still a matter of debate \citep{VazquezSemadeniEtAl2011, KoertgenBanerjee2015, KoertgenEtAl2018, NtormousiHennebelle2019}. The complex interplay between the magnetic field and the dynamics in the gas includes the strength of the field, its geometry as well as complicated gas motions, which can result in efficient field amplification via the magnetic dynamo \citep[e.g.][]{Brandenburg2018,Subramanian2019}.

The ordered component of the field on galactic scales is observed with field strengths of $1-10\,\mu\mathrm{G}$ \citep{Beck2009, FletcherEtAl2011}. In the interstellar medium the field strength can reach significantly higher values of several $100\,\mu\mathrm{G}$ in particular in dense gas like molecular clouds and star forming regions (see review by \citealt{Crutcher2012}).

The magnetic field strength is expected to be connected to the density of the gas $B\propto n^{\kappa}$ and the gas velocities \citep[e.g.][]{Mestel1966,ChandrasekharFermi1953,MouschoviasCiolek1999}. In the ideal MHD approximation a gas flow perpendicular to the the field orientation compresses the field and increases its strength linearly with the density ($\kappa=1$). Contrary, gas motions parallel to the field lines do not influence the field intensity, i.e. $\kappa=0$. Consequently, the orientation of the field lines with respect to the gas flow is an important quantity.

The orientation of the gas velocities and the resulting gas structures with respect to the field lines can be intuitively pictured in two extreme cases. In low-density regions with weak magnetic fields, the density and magnetic field structures are primarily determined by the gas flow. Motions compress the gas together with the field and create configurations of field lines parallel to the gas structures, i.e. with resulting density gradients perpendicular to $\vektor{B}$. One prominent example is the hot Local Bubble around the solar system. \citet{AlvesEtAl2018} find that the local magnetic field does not follow the large-scale field of the Milky Way. Instead the field is distorted by the hot expanding gas in the Local Bubble, as found by \citet{PelgrimsEtAl2020} using dust polarization data. The other extreme are dense regions with strong fields, where the flow is directed along the field lines. The resulting structures and filaments are oriented perpendicular to $\vektor{B}$, i.e. with density gradients parallel to the magnetic field vector.

Simulations of the turbulent ISM with $\langle n\rangle=5\times10^2\,\percc$, ($\rho\sim10^{-21}\,\gpercc$) in $4\,\mathrm{pc}$ periodic boxes with decaying turbulence and varying thermal to magnetic energy ratios confirm this hypothesis \citep{SolerEtAl2013}. The converging flow simulations of \citet{ChenKingLi2016} include self-gravity, and a perturbed turbulent velocity field in a 1\,pc box. Using isothermal gas at $10\,\mathrm{K}$ at a mean density of $10^3\,\percc$ ($\rho\sim2\times10^{-21}\,\gpercc$) they identify a threshold density of $10^5\,\percc$ ($\rho\sim2\times10^{-19}\,\gpercc$) at which the gradient of the density aligns with the magnetic field. They connect this transition with the onset of gravitational instability. Analysing the MHD equations \citet{SolerHennebelle2017} show that the two extreme cases of alignment of the field with the density gradient represent equilibrium points towards which the systems tend to evolve. Observational evidence was presented by \citet{PlanckXXXV2016} and \citet{SolerEtAl2017}, investigating the magnetic field in the plane of the sky and the structures in the column density. More recent observations \citep{AlinaEtAl2019, FisselEtAl2019} demarcate the transition between the two alignment regimes at a number density of $n\sim10^{3}\,\percc$ ($\rho\sim2\times10^{-21}\,\gpercc$). \citet{AlinaEtAl2019} distinguish between filaments embedded in low and high-density environments. In the former case the fields in the filament show a predominantly parallel orientation with respect to the background. For filaments embedded in an environment above $\Sigma\approx0.003\,\gperscm$ ($N\approx10^{21}\,\perscm$) the orientation changes from parallel to perpendicular. However, this behaviour is not universal for all structures. In clumps \citet{AlinaEtAl2019} find both perpendicular and parallel relative orientations regardless of the environmental column density. Contrary to the well confirmed orientation behaviour in three-dimensional simulations, the corresponding measurements in projections show less strong signals and can differ between different objects and different lines of sight. Variations in the numerical experiments modify the reliability of the projected measurements.

Using high-resolution simulations of the magnetized SN-driven interstellar medium from spatial scales of $500\,\mathrm{pc}$ down to $0.25\,\mathrm{pc}$ we self-consistently follow the formation of structures from the hot phase of the ISM to dense star forming regions together with the evolution of interstellar turbulence and the magnetic field. The simulation set-up has been tested in previous studies within the SILCC project \citep{WalchEtAl2015, GirichidisEtAl2016b} including supernova-driven turbulence and a non-equilibrium chemical network. Bridging the spatial scales from $500\,\mathrm{pc}$ down to $0.1\,\mathrm{pc}$ while accurately following the dynamics has been tested by \citet{SeifriedEtAl2017} and applied to simulations of star clusters \citep{HaidEtAl2018, HaidEtAl2019}. A focus on the magnetic fields with their general impact on the formation of dense and molecular gas has been presented by \citet{PardiEtAl2017} and \citet{GirichidisEtAl2018b}. In this study we extend the previous models by increasing the numerical resolution and the statistical sample in dense regions. \citet{PardiEtAl2017} and \citet{GirichidisEtAl2018b} reach resolutions of up to $1\,\mathrm{pc}$. \citet{SeifriedEtAl2017} resolve the dense gas down to scales of approximately $0.1\,\mathrm{pc}$ for two selected molecular clouds. In the present study we reach resolutions up to $0.25\,\mathrm{pc}$ but do not restrict this high resolution to one or two selected clouds. Instead all dense clouds ($\sim500$) in the simulation box are resolved with the highest resolution, which increases the statistics in the dense gas. We confirm the naturally driven alignment of gas gradients and gas motions with respect to the magnetic field lines across a large dynamic range from the hot ionized gas to the cold molecular phase with self-consistently generated SN-driven motions. We investigate in detail the change in orientation from intermediate to high densities, in which gravitational instability occurs. A special focus of this study is the connection between the three-dimensional effects of the local hydrodynamical quantities and the resulting projection effects that are conceptually comparable to observational column density structures and plane-of-the-sky magnetic fields. The complications of small scale structures and a resulting averaging effect are presented. We highlight how the structure of the turbulence could weaken the observational signal for an alignment.

The paper is structured as follows. In Section~\ref{sec:methods} we give a brief introduction to the numerical methods, the included physical ingredients and the initial conditions of the simulations. Section~\ref{sec:morphology} covers an overview of the global evolution. The orientation of the magnetic field with respect to other hydrodynamical quantities is discussed in Section~\ref{sec:B-orientation}. Here we focus on the three-dimensional data. The resulting projection effects are investigated in Section~\ref{sec:coldens-effects}. The results are discussed and compared to other work in Section~\ref{sec:discussion} before we conclude in Section~\ref{sec:conclusion}.

\section{Methods and Simulations}
\label{sec:methods}

The simulations are based on the SILCC set-up (SImulating the Life Cycle of molecular Clouds, \citealt{WalchEtAl2015, GirichidisEtAl2016b}). The physics modules and the set-up are described in detail in \citet{WalchEtAl2015}. The effects of magnetic fields have been investigated in \citet{GirichidisEtAl2018b} with the same set-up but lower resolution. We therefore only summarize the main features briefly here.

We simulate a stratified box with a size of $0.5\times0.5\times0.5\,\mathrm{kpc}^3$ using periodic boundary condition along $x$ and $y$ and diode boundary conditions along the stratification axis $z$, which means the gas can leave but not enter the simulation box. Consequently, outflowing gas that passes the $z$ boundary cannot return as a fountain flow, which however, takes perceptibly longer than the formation of molecular clouds \citep[see e.g.][]{HillEtAl2012, GirichidisEtAl2016b, KimOstriker2018}. We solve the equations of ideal MHD using the HLLR5 solver \citep{Bouchut2007, Bouchut2010, Waagan2009, Waagan2011} in the adaptive mesh refinement code \textsc{Flash} in Version 4 (\citealt{FLASH00,DubeyEtAl2008}, \url{http://flash.uchicago.edu/site/}).

Heating and radiative cooling are connected to a chemical network that follows the non-equilibrium abundances of ionized (H$^+$), atomic (H) and molecular (H$_2$) hydrogen as well as singly ionized carbon (C$^+$) and carbon monoxide (CO). The hydrogen chemistry is based on \citet{GloverMacLow2007a, GloverMacLow2007b} as shown in \citet{MicicEtAl2012} and is extended by the CO model by \citet{NelsonLanger1997}. We follow the formation and destruction of H$^+$ including collisional ionization, ionization by cosmic rays and X-rays and radiative recombination. The formation of H$_2$ follows \citet{Hollenbach89} and includes the destruction by cosmic ray ionization, by collisional dissociation in hot gas and by photodissociation in the presence of the interstellar radiation field. The radiative cooling covers contributions from the fine structure lines of C$^+$, O and Si$^+$. In addition, rotational and vibrational lines of H$_2$ and CO, Lyman-$\alpha$ cooling as well as the transfer of energy from the gas to the dust \citep{GloverEtAl2010, GloverClark2012b} are taken into account. At temperatures above $10^4\,\mathrm{K}$ we assume collisional ionization equilibrium and use cooling rates from \citet{GnatFerland2012}. Hydrogen is not assumed to be in collisional ionization equilibrium, so its contribution to the cooling is computed self-consistently using the actively followed abundances.

Heating of the gas includes several processes. We include spatially clustered supernovae (SNe), which are described in more detail below. In addition, cosmic ray \citep{GoldsmithLanger1978} and X-ray heating \citep{WolfireEtAl1995} are included, which are spatially and temporally constant and are not correlated with the SNe. We use a CR ionization rate of $\zeta_\mathrm{CR}=3\times10^{-17}\,\mathrm{s}^{-1}$ and a corresponding heating rate per unit volume of $\Gamma_\mathrm{CR}=3.2\times10^{-11}\zeta_\mathrm{CR}\,n\,\mathrm{erg\,s}^{-1}\,\mathrm{cm}^{-3}$. Photoelectric heating follows \citep{BakesTielens1994, Bergin2004, WolfireEtAl2003}. The constant interstellar radiation field with a strength of $1.7$ in the units of the Habing field $G_0$ \citep{Habing1968, Draine1978} is locally attenuated in dense shielded regions. The local column depth is computed using the TreeCol algorithm \citep{ClarkGloverKlessen2012}, which has been implemented and optimized for \textsc{Flash} as described in \citet{WuenschEtAl2018}. The dust-to-gas mass ratio is set to 0.01 with dust opacities based on \citet{MathisMezgerPanagia1983} and \citet{OssenkopfHenning1994}.

The initial gas distribution follows a Gaussian profile with a scale height of $30\,\mathrm{pc}$ and a central density of $\rho_0=9\times10^{-24}\,\mathrm{g\,cm}^{-3}$, which yields a total gas column density of $\Sigma=10\,\mathrm{M}_\odot\mathrm{pc}^{-2}$. The temperature is adjusted such that the gas is initially in pressure equilibrium. This corresponds to a central temperature of $4600\,\mathrm{K}$ and a value of $T=4\times10^8\,\mathrm{K}$ at large altitudes, where we apply a lower floor to the density ($\rho_\mathrm{min}=10^{-28}\,\mathrm{g\,cm}^{-3}$).

Besides self-gravity we include an external potential that accounts for the stellar component of the disc. We use an isothermal sheet \citep{Spitzer1942} with a stellar gas surface density of $30\,\mathrm{M}_\odot\mathrm{pc}^{-2}$ and a vertical scale height of $100\,\mathrm{pc}$. The gravitational forces are computed using the tree-based method by \citet{WuenschEtAl2018}.

The initial magnetic field is oriented along the $x$ direction. The central field strength in the midplane at $z=0$ is set to $B_0=3\,\mu\mathrm{G}$ and scales vertically with the density, $B_x(z)=B_{x,0}\,[\rho(z)/\rho(z=0)]^{1/2}$. We do not introduce an initial small scale tangled perturbation of the field. We note that the initial magnetic field strengths is not strong enough to support the disc against gravitational collapse, see Appendix~\ref{sec:app-mass-to-flux-ratio}.

Stellar feedback is included as a constant SN rate. We use the Kennicutt-Schmidt relation \citep{KennicuttSchmidt1998} to find a star formation rate surface density and convert it to a SN rate taking the initial stellar mass function from \citet{Chabrier2003}. For each of the $15$ explosions per Myr in our box we inject $10^{51}\,\mathrm{erg}$ of thermal energy. We apply the SN feedback from the beginning of the simulation and distinguish between different types and distributions \citep{TammannLoefflerSchroeder1994}. We include a type~Ia component (20\% of the SNe) with a uniform distribution in $x$ and $y$ and a Gaussian distribution in $z$ with a scale height of $300\,\mathrm{pc}$ \citep{BahcallSoneira1980, Heiles1987}. The remaining 80\% are of type~II with a vertical scale height of $90\,\mathrm{pc}$ and are further split into a runaway component (60\%) of individual explosions and a clustered fraction (40\%), which are spatially grouped with numbers ranging from 7 to 40 SNe per cluster \citep{Heiles1987, KennicuttEtAl1989, McKeeWilliams1997, ClarkeOey2002}.

We start with an initial resolution of $128^3$ cells and refine based on the density \citep{Loehner1987} as well as on the molecular gas fraction as in \citet{GirichidisEtAl2018b}. For the first $19.7\,\mathrm{Myr}$ we run the simulation using a maximum effective resolution of $512^3$ ($\Delta x\approx1\,\mathrm{pc}$). We then successively add two additional levels of refinement as presented in \citet{SeifriedEtAl2017}, each increasing the effective resolution by a factor of 2. From $19.7-20.6\,\mathrm{Myr}$ we keep the effective resolution at $1024^3$ ($\Delta x\approx0.5\,\mathrm{pc}$) before we switch to $2048^3$ ($\Delta x\approx0.25\,\mathrm{pc}$) and continue the simulation for another $\sim200\,\mathrm{kyr}$. The $1024^3$ simulation is continued up to a final time of $20.9\,\mathrm{Myr}$.

\begin{table*}
\begin{minipage}{12cm}
    \caption{List of simulations}
    \label{tab:simulations}
    \begin{tabular}{lcccccccccccc}
      Name   & $B_0$ & eff. res. & $\Delta x$ & $t_\mathrm{start}$ & $t_\mathrm{stop}$ & ana. time & \#sim. files & avg. \#cells\\
      & ($\mu$G) &      & (pc)  &  Myr & Myr & Myr & & & \\
      \hline
      \texttt{B3-2pc}    & $3$   & $256^3$  & $1.95$   & \phantom{1}0.0 & 60.0 & 20.0-21.0 & 11 & $6.6\times10^6$\\
      \texttt{B3-1pc}    & $3$   & $512^3$  & $0.98$   & \phantom{1}0.0 & 60.0 & 20.0-21.0 & 11 & $3.4\times10^7$\\
      \texttt{B3-0.5pc}  & $3$   & $1024^3$ & $0.49$   & 19.7           & 20.9 & 20.4-20.9 & 11 & $2.0\times10^8$\\
      \texttt{B3-0.25pc} & $3$   & $2048^3$ & $0.245$  & 20.6           & 20.8 & 20.7-20.8 & 16 & $3.3\times10^8$\\
      \hline
    \end{tabular}

    \medskip
    Presented are the name of the simulation, the initial magnetic field strength in the midplane, the effective numerical resolution, the corresponding spatial resolution of the smallest cell, the starting time for the simulation with the given resolution, the maximum simulation time, the resulting analysis time span, the total number of files that we use for averaging over time as well as the total number of cells in the simulation box averaged over the analysis time span.
\end{minipage}
\end{table*}

Besides the high resolution simulation we also include simulations \texttt{B3-2pc}, \texttt{B3-1pc} from \citet{GirichidisEtAl2018b} in the analysis, which allows us to compare to lower resolutions. An overview of the simulations used is presented in table~\ref{tab:simulations}. We emphasize that we do not start the analysis right after switching to a higher refinement level in order to allow for an adjustment of the hydrodynamical quantities to the higher resolution. This adjustment also includes a population of turbulent modes and magnetic entanglement. We follow the recommendations by \citet{SeifriedEtAl2017} who use a very similar set-up and code and suggest 200 hydrodynamical time steps to allow the quantities to adjust to the new resolution.


\section{Morphological evolution and global analysis}
\label{sec:morphology}

\begin{figure*}
    \centering
    \includegraphics[width=\textwidth]{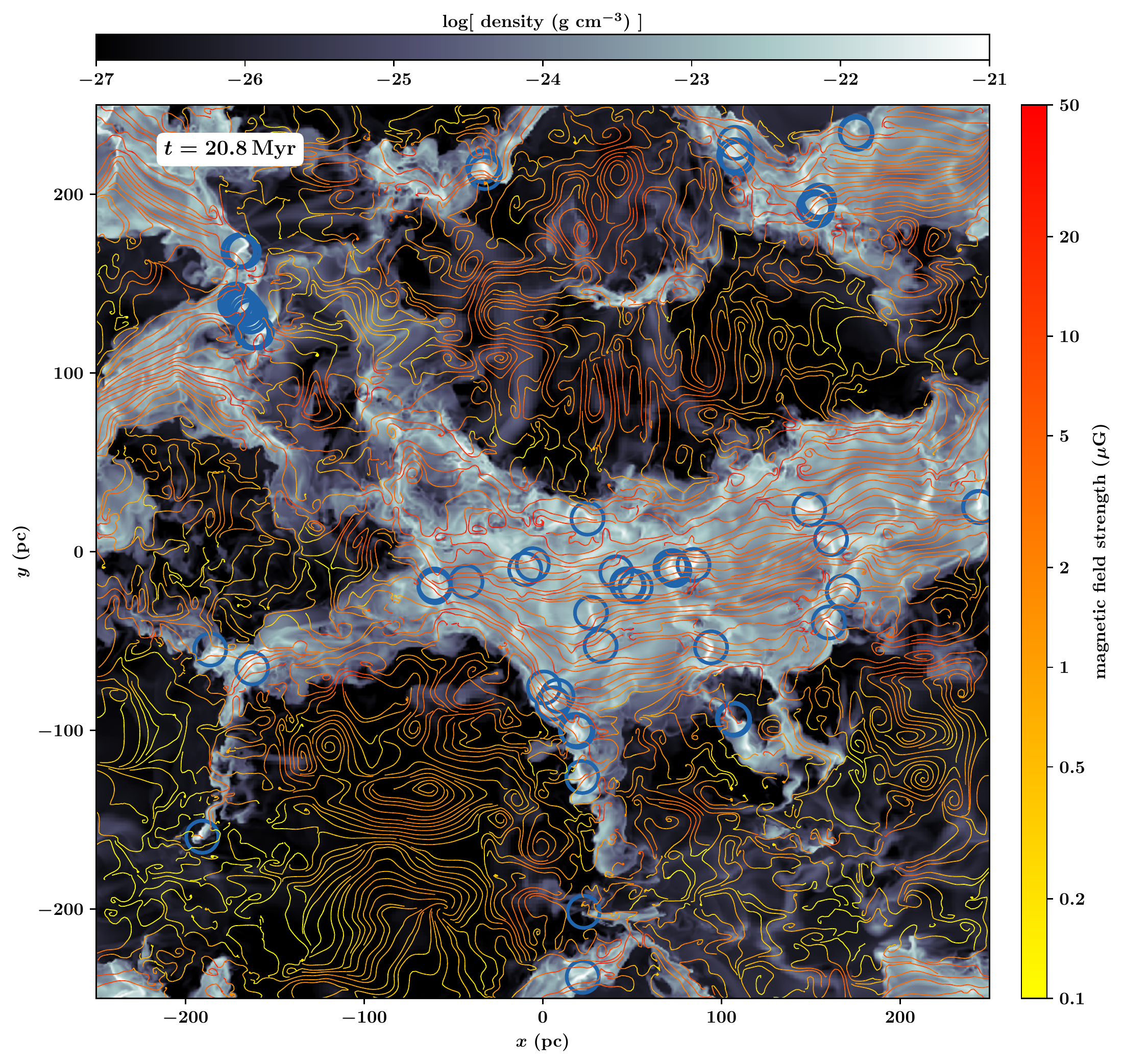}
    \caption{Cut through the centre of the box at $z=0$ of simulation \texttt{B3-0.25pc} at $t=20.8\,\mathrm{Myr}$. The density is colour coded in gray-scale. The magnetic field lines are overplotted with the field strength indicated by the streamline colour. The blue circles indicate the identified clouds based on local minima of the gravitational potential.}
    \label{fig:dens-mag-streamlines}
\end{figure*}

\begin{figure}
    \centering
    \includegraphics[width=8cm]{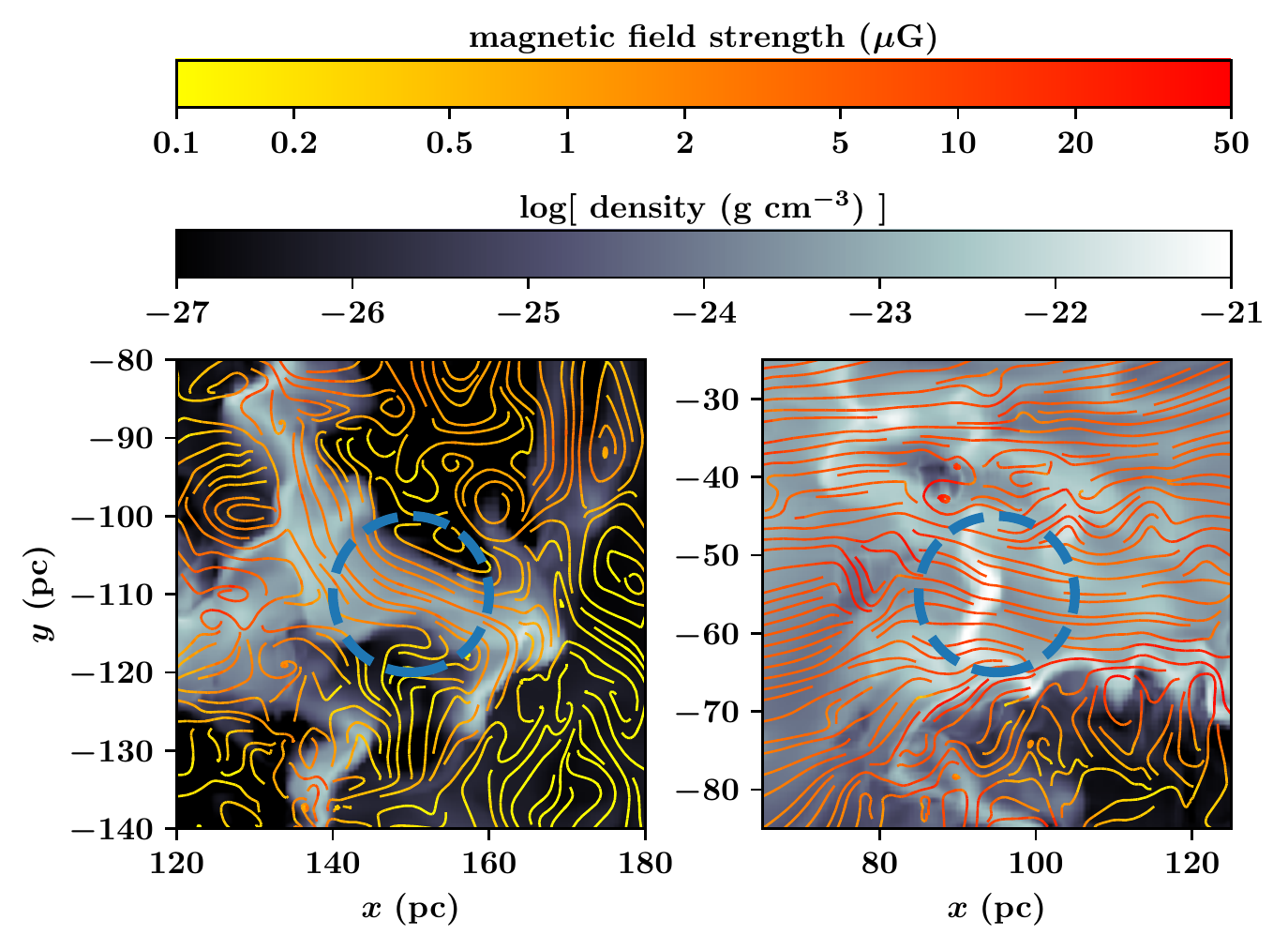}
    \caption{Two zoom-in regions to illustrate the two extremes of the magnetic field orientation with respect to the gas density. In the left-hand picture the magnetic field is aligned with the curved filament ($\vektor{B}\perp\nabla\rho$). In the right-hand counterpart the dense vertical filament formed via accretion along the horizontal field lines ($\vektor{B}\parallel\nabla\rho$). For comparison we also add circles corresponding to the size of the analysis volume.}
    \label{fig:dens-mag-streamlines-zoom}
\end{figure}

\begin{figure}
    \centering
    \includegraphics[width=8cm]{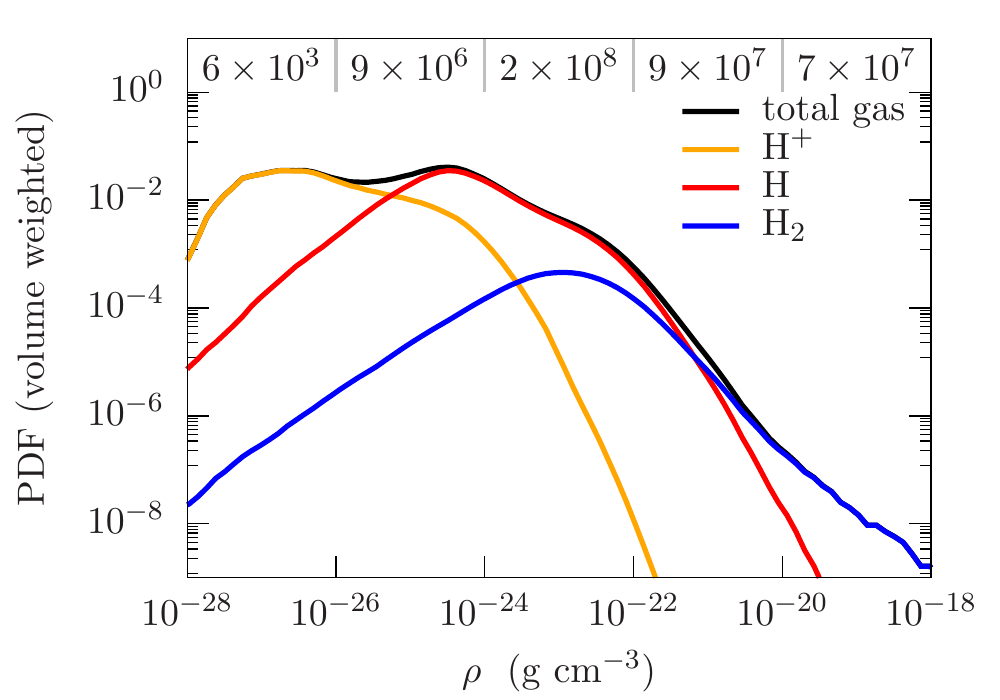}
    \caption{Density PDF split into the three hydrogen components for simulation \texttt{B3-0.25pc} at $t=20.8\,\mathrm{Myr}$. Above a density of $\rho\sim10^{-20}\,\mathrm{g\,cm}^{-3}$ ($n\sim4000\,\percc$) a powerlaw tail has developed, which is dominated by molecular hydrogen and indicates gravitational collapse.}
    \label{fig:density-pdf-chem-composition}
\end{figure}

The SNe start exploding from the beginning of the simulation and immediately begin to introduce structures in the gas. After $\sim~10-15\,\mathrm{Myr}$ the explosions have formed voids and filaments that continue to contract because of SN-driven dynamics as well as gravitational attraction in the overdense regions. The density range spans around ten orders of magnitude at the end of the simulation. The magnetic field has been compressed and tangled due to the SN-driven motions and self-gravity. We illustrate the structure of the density and the magnetic field in Fig.~\ref{fig:dens-mag-streamlines} in a cut through the midplane at $z=0$ for simulation \texttt{B3-0.25pc} at the end of the simulation at $t=20.8\,\mathrm{Myr}$. Colour coded in the background is the gas density. The magnetic field lines are overplotted with the field strength indicated by the streamline colour. The circles indicate clouds identified by local minima of the gravitational potential, see \citet{GirichidisEtAl2018b} for more details and a comparison to other methods. We only show circles for identified clouds whose centres lie within $z=\pm3\,\mathrm{pc}$ in order to illustrate the correlation with the density cut through $z=0$, which explains why not all dense regions are marked. The first molecular clouds have formed with peak densities of $\rho\gtrsim10^{-18}\,\mathrm{g\,cm}^{-3}$ ($n\gtrsim4\times10^5\,\percc$). Two zoom-in regions are shown in Fig.~\ref{fig:dens-mag-streamlines-zoom} to illustrate the two extreme cases of the orientation of the field lines with respect to the gas structures.

We show the probability density function (PDF) for the total density as well as the different chemical states in Fig.~\ref{fig:density-pdf-chem-composition}. Up to a density of $\rho\sim10^{-25}\,\mathrm{g\,cm}^{-3}$ ($n\sim0.04\,\percc$) the gas is dominated by ionized hydrogen. The gas up to $\rho\sim10^{-21}\,\mathrm{g\,cm}^{-3}$ ($n\sim400\,\percc$) mainly consists of atomic gas. Above that density the gas is mainly in molecular form and a powerlaw tail has developed, which is an indicator for gravitational collapse \citep{Klessen2000, SlyzEtAl2005, GirichidisEtAl2014}. We also show the number of computational cells in each density range above the curves to ensure that the following analysis is based on a statistically relevant sample of cells.


\section{Orientation of the magnetic field}
\label{sec:B-orientation}

\subsection{Distribution of angles}

Before we analyse the simulation data, we would like to highlight the properties of a statistical distribution of angles. Let us consider randomly distributed vectors with unit length. For randomly oriented vectors in a two-dimensional plane the \emph{angle} $\theta_\mathrm{2D}$ between them is uniformly distributed. In three dimensions a random orientation of vectors -- corresponding to a uniform point distribution on the surface of the unit sphere -- results in a uniform distribution of the \emph{cosine of the angle}, $\cos\theta$, where we omit the subscript \emph{3D}.

\begin{figure*}
    \centering
    \includegraphics[width=0.45\textwidth]{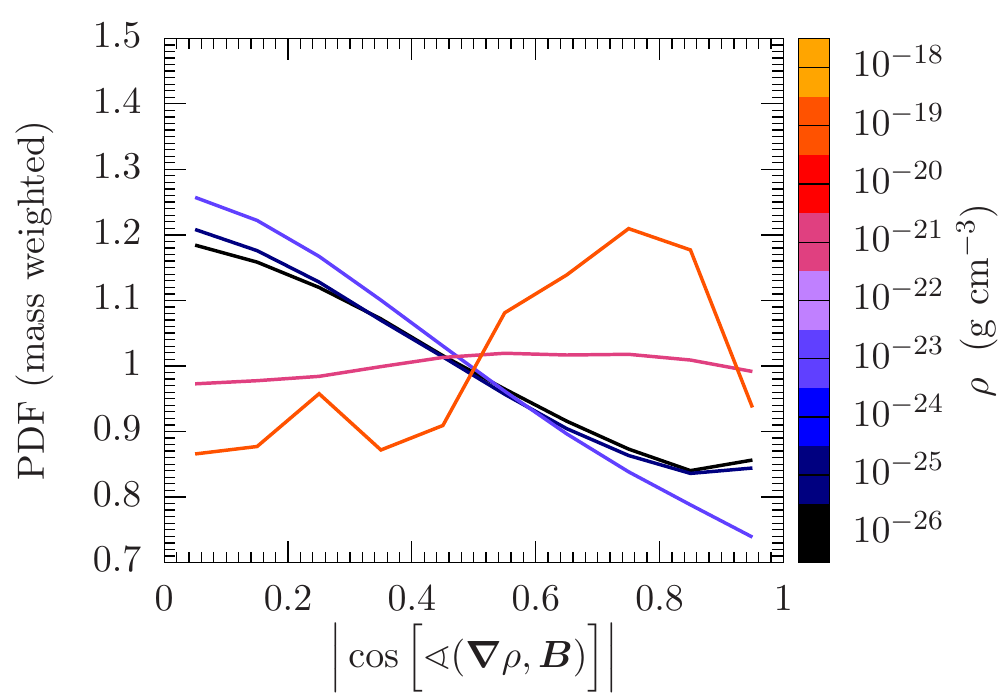}
    \includegraphics[width=0.45\textwidth]{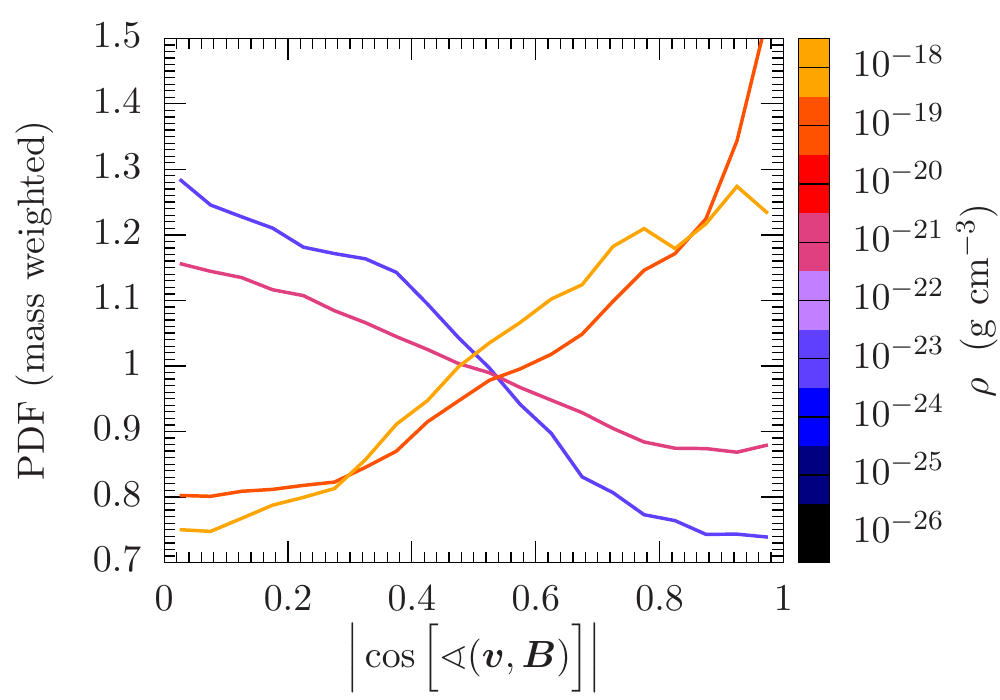}
    \caption{Distribution of $|\cos\theta_\subrhoB|$ (left-hand panel) and $|\cos\theta_\subvelB|$ (right-hand panel) for simulation \texttt{B3-0.25pc} at $t=20.77\,\mathrm{Myr}$. We show distributions for different density ranges. The curves indicate that for low densities both the gradient of the density as well as the gas flow are preferentially oriented perpendicular to the magnetic field. Above densities of $\rho\sim10^{-20}\,\mathrm{g\,cm}^{-3}$ ($n\sim4000\,\percc$) both quantities tend to be parallel to the field.}
    \label{fig:angle-gradrhoB-PDF}
\end{figure*}

We investigate the distribution of the angle between the gradient of the density and the magnetic field vector,
\begin{equation}
    \cos\theta_\subrhoB = \cos[\sphericalangle(\vnabla\rho, \vektor{B})] = \frac{\vnabla\rho\cdot\vektor{B}}{||\vnabla\rho||\cdot||\vektor{B}||}
\end{equation}
as well as the angle between the velocity of the gas and $\vektor{B}$,
\begin{equation}
    \cos\theta_\subvelB = \cos[\sphericalangle(\vektor{v}, \vektor{B})] = \frac{\vektor{v}\cdot\vektor{B}}{||\vektor{v}||\cdot||\vektor{B}||}.
\end{equation}
Since flipping the direction of magnetic field everywhere ($\vektor{B}\rightarrow-\vektor{B}$) does not influence the ideal MHD dynamics, it is sufficient to analyse $|\cos\theta|$ in both cases. The extreme case of $|\cos\theta_\subrhoB|\approx 1$, i.e. parallel vectors, is illustrated in Fig.~\ref{fig:dens-mag-streamlines-zoom} in the left-hand panel inside the circle. The opposite, $|\cos\theta_\subrhoB|\approx 0$, for perpendicular vectors is shown in the right-hand panel, respectively.

For the angle between $\vektor{v}$ and $\vektor{B}$, we are interested in the velocities in the local rest frame, i.e., we would like to subtract bulk motions. This is particularly important for dense molecular structures where the bulk motion of the cloud can exceed the low internal velocity dispersion. We apply the procedure described in detail in \citet{GirichidisEtAl2018b}, which can be summarised as follows. We find the local minima of the gravitational potential and use that location as the centre of the cloud. A spherical volume with a radius of $10\,\mathrm{pc}$ around that minimum is used as the analysis volume. We subtract the centre-of-mass velocity and measure the distribution of $\theta_\subvelB$. For the binning in $\rho$ we use the density at the centre of the cloud (rather than the average density of the analysis volume) in order to avoid a strong sensitivity on the analysis radius. We investigated this using different radii from 2 to 15 pc and found very little deviation in the obtained results. We note that there can be multiple local minima of the gravitational potential within a radius of $10\,\mathrm{pc}$. Because the clouds are identified based on the minima, densities below $\rho\sim10^{-24}\,\gpercc$ are not included in the analysis of $\theta_\subvelB$.

Figure~\ref{fig:angle-gradrhoB-PDF} shows the mass weighted PDF of $|\cos\theta_\subrhoB|$ (left-hand panel) and $|\cos\theta_\subvelB|$ (right-hand panel) for different ranges of the gas density for run \texttt{B3-0.25pc} at the end of the simulation. The distributions look very similar for ionized, atomic and warm molecular gas. The density, at which the distributions start to change all correspond to cold molecular gas. For $\rho\lesssim10^{-22}\,\mathrm{g\,cm}^{-3}$ ($n\sim40\,\percc$) the distributions peak at $|\cos\theta|=0$, i.e. there is a preference of the magnetic field to be oriented perpendicular to the density gradient as well as to the velocity vector (this correponds to the magnetic field being aligned with the filament, as in the left panel of Fig.~\ref{fig:dens-mag-streamlines-zoom}). The transition occurs at approximately $\rho\sim10^{-21}\,\gpercc$ ($n\sim400\,\percc$. For $\rho\gtrsim10^{-20}\,\mathrm{g\,cm}^{-3}$ the gradient of the density and the velocities tend to be more aligned with the magnetic field vector as in the example shown in the right-hand panel of Fig.~\ref{fig:dens-mag-streamlines-zoom}. We also note that the transition between the two regimes occurs at slightly different densities for $\theta_\subrhoB$ and $\theta_\subvelB$, which is, however, not significant (see analysis of $\xi_\mathrm{3D}$ and Fig.~\ref{fig:xi-3D} in the next Section). The differences in the mass and volume weighted distributions are minor (not shown). The two extreme configurations in the orientation are not very prominent in each cell and are difficult to visually spot in the simulations, i.e. the two regions shown in Fig~\ref{fig:dens-mag-streamlines-zoom} are particularly clean examples. None the less there is a clear signal for a primarily perpendicular orientation between $\vnabla\rho$ and $\vektor{B}$ as well as between $\vektor{B}$ and $\vektor{v}$.

\subsection{Quantifying the distribution}

\begin{figure}
    \centering
    \includegraphics[width=8cm]{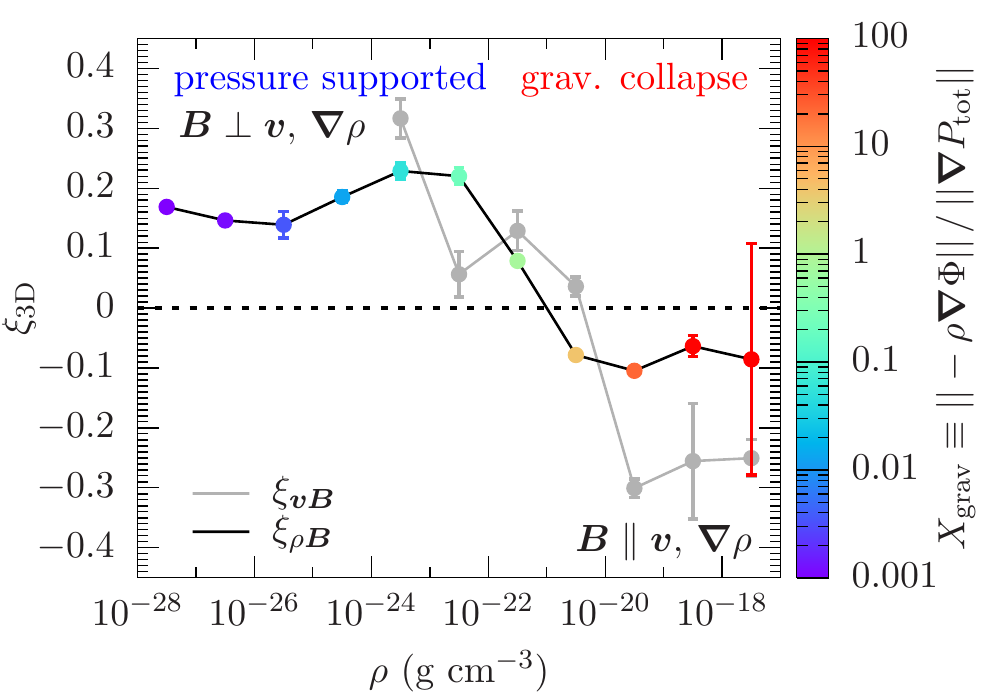}
    \caption{Three-dimensional measure $\xi_\mathrm{3D}$ as a function of density for simulation \texttt{B3-0.25pc} averaged over the entire analysis time. We note a clear switch from positive to negative values at $\rho\sim10^{-21}\,\mathrm{g\,cm}^{-3}$ ($n\sim400\,\percc$). The low density gas is pressure supported ($X_\mathrm{grav}<1$) with a preference for the magnetic field to be perpendicular to the density gradient. Dense regions are dominated by the gravitational potential ($X_\mathrm{grav}>1$) and show an alignment of $\vektor{B}$ with the gradient of the density. The temporal variations are small for most of the data points except for the highest density due to the lower number of cells. We do not show $\xi_\subvelB$ for $\rho<10^{-24}\,\mathrm{g\,cm}^{-3}$, see text.}
    \label{fig:xi-3D}
\end{figure}

\begin{figure}
    \centering
    \includegraphics[width=8cm]{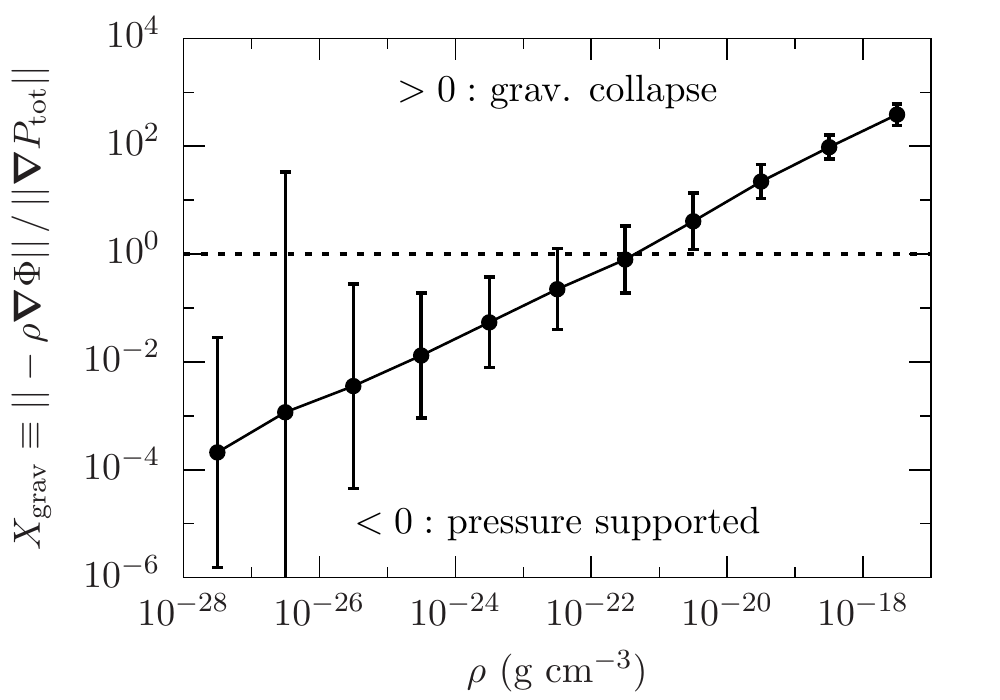}
    \caption{Ratio of gravitational attraction over pressure support (see Eq.~\ref{eq:X_grav}) as a function of density for simulation \texttt{B3-0.25pc} averaged over the entire analysis time. The ordinate corresponds to the colour coded values in Fig.~\ref{fig:xi-3D} showing the mean over all cells within the given density regime. The errorbars indicate the standard deviation. At low densities the variations in the ratio are large. With increasing density the standard deviation reduces to negligible values. Accounting for the uncertainty in $X_\mathrm{grav}$, the transition density ranges from $\sim3\times10^{-23}$ to $\sim3\times10^{-21}\,\gpercc$ ($n\sim10-1000\,\percc$).}
    \label{fig:grav-ratio}
\end{figure}

We quantify the shape of the distributions using the \emph{histogram of relative orientations} (HRO) \citep{SolerEtAl2013,PlanckXXXV2016}. In its two-dimensional form it measures the distribution of angles between the iso-contour of the column density ($\Sigma$) and the magnetic field in the plane of the sky in the range from $-90^\circ$ to $90^\circ$. The quantity $\xi$ \citep{PlanckXXXV2016} is defined as
\begin{equation}
    \xi_\mathrm{2D} = \frac{A_\mathrm{c}-A_\mathrm{e}}{A_\mathrm{c}+A_\mathrm{e}},
    \label{eq:xi_2D}
\end{equation}
where $A_\mathrm{c}$ and $A_\mathrm{e}$ are integrated fractions of the histogram in the range $-22.5^\circ<=\phi<=22.5^\circ$ and $|\phi|>67.5^\circ$. These correspond to the two quarters of low and high modulus of the angles. We apply a similar analysis to our 3D data which is given as a probability distribution function, $H_{\cos\theta}$. Here
$H_{\cos\theta}$ can refer to the distribution in either $\cos \theta_\subvelB$ or $\cos \theta_\subrhoB$. We again note that we only need to consider the modulus $|\cos\theta|$ because the direction of the field line does not enter the dynamics in ideal MHD, i.e. an orientation parallel to the field line is equivalent to an antiparallel configuration, so $H_{\cos\theta}\rightarrow H_{|\cos\theta|}$. We thus arrive at a comparable quantity in three dimensions,
\begin{equation}
    \xi_\mathrm{3D} = \frac{A'_\mathrm{c}-A'_\mathrm{e}}{A'_\mathrm{c}+A'_\mathrm{e}},
\end{equation}
with $A'_\mathrm{c}$ and $A'_\mathrm{e}$ being the corresponding integrated quarters of the left and right-hand side of the $H_{|\cos\theta|}$ histogram,
\begin{align}
    A'_\mathrm{c} &= \int_0^{1/4}H_{|\cos\theta|}\,\mathrm{d}\cos\theta,\\
    A'_\mathrm{e} &= \int_{3/4}^{1}H_{|\cos\theta|}\,\mathrm{d}\cos\theta.
\end{align}
Values of $\xi_\mathrm{3D}>0$ indicate that $|\cos\theta|$ is preferentially close to 0 ($\theta\sim90^\circ$), so $\vektor{B}$ is perpendicular to $\vnabla\rho$ and to $\vektor{v}$. This corresponds in 2D to $\vnabla\Sigma$ being perpendicular to $\vektor{B}$, i.e. iso-contours of $\Sigma$ being parallel to $\vektor{B}$. We would like to point out that \citet{SolerEtAl2013} also include a study on the three-dimensional data, although not with such a quantitative figure of merit for the histogram shapes.

A more precise analysis of the transition from $\xi_\mathrm{3D}>0$ to $\xi_\mathrm{3D}<0$ is shown for both angles in Fig.~\ref{fig:xi-3D} as a function of density for run \texttt{B3-0.25pc}. We discuss the effects of resolution in appendix~\ref{sec:app-resolution}. Again, for $\theta_\subvelB$ we first identify clouds based on local minima of the gravitational potential in order to subtract the bulk motion of the cloud, which excludes densities below $\rho\lesssim10^{-24}\,\mathrm{g\,cm}^{-3}$ ($n\lesssim0.4\,\percc$). The points are averaged over 16 data snapshots spanning a total evolutionary time of $80\,\mathrm{kyr}$. The errorbars indicate the standard deviation of temporal variations. Colour coded is the deviation from hydrostatic equilibrium encoded in the ratio
\begin{equation}
    X_\mathrm{grav} \equiv \frac{||-\rho\vnabla\Phi||}{||\vnabla P_\mathrm{tot}||},
    \label{eq:X_grav}
\end{equation}
where $\Phi$ is the gravitational potential and $P_\mathrm{tot}=P_\mathrm{th}+P_\mathrm{mag}$. Here $P_\mathrm{th}=\rho k_\mathrm{B}T/(\mu m_\mathrm{H})$ and $P_\mathrm{mag}=B^2/(8\pi)$ are the thermal and magnetic pressure with the Botzmann constant $k_\mathrm{B}$, the temperature $T$, the mean molecular weight $\mu$, and the hydrogen mass $m_\mathrm{H}$. For $X_\mathrm{grav} > 1$ the gravitational acceleration dominates over the stabilizing pressure. Values below unity indicate stability. The colors in the plot show the mean over all cells in the given density bin. The standard deviation for $X_\mathrm{grav}$ is shown in Fig.~\ref{fig:grav-ratio}. For densities below $10^{-22}\,\mathrm{g\,cm}^{-3}$ $\xi_\mathrm{3D}$ is positive for both angles with values ranging from $\xi_\subrhoB=0.15-0.25$ and $\xi_\subvelB=0.05-0.3$, i.e. $\vektor{v}$ and $\vnabla\rho$ are preferentially perpendicular to $\vektor{B}$. In this density range the pressure can support the gas against gravitational contraction. Between $10^{-21}\,\mathrm{g\,cm}^{-3}$ and $10^{-20}\,\mathrm{g\,cm}^{-3}$ there is a steep transition to negative values around $\xi_\subrhoB\sim-0.1$ and $\xi_\subvelB\sim-0.25$. The errorbars are small for all but the highest density. However, the total number of cells at the highest density is small, so this data point should not be overinterpreted. The change in orientation is accompanied by a transition from pressure supported to gravitationally collapsing gas, see Fig.~\ref{fig:grav-ratio}. Within the statistical uncertainty of the standard deviation the transition density between pressure supported and gravity dominated regions spans two orders of magnitude from $\sim3\times10^{-23}$ to $\sim3\times10^{-21}\,\gpercc$ ($n\sim10-1000\,\percc$) with the transition of the mean at $\sim3\times10^{-22}$ ($n\sim100\,\percc$). This is further justified in Fig.~\ref{fig:angle-acc-Ptot-PDF}, see below.

\subsection{The dense gas is gravitationally unstable}

\begin{figure}
    \centering
    \includegraphics[width=8cm]{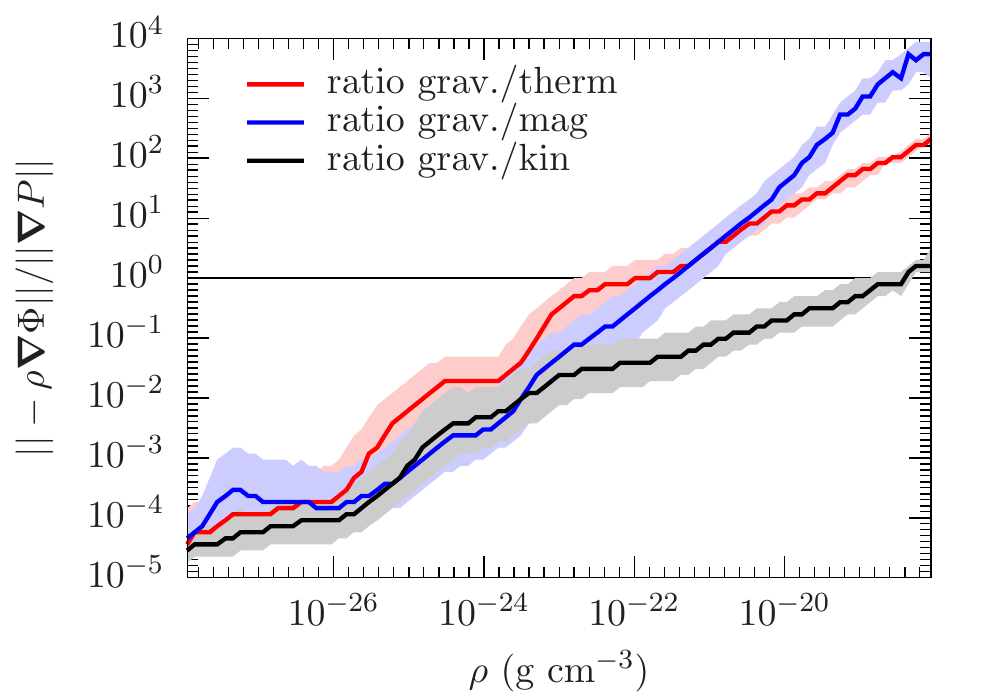}
    \caption{Ratio of the absolute value of the gravitational force per unit volume to the absolute value of the pressure gradients as a function of density. The curves show the median of the distribution for run \texttt{B3-0.25pc} at the end of the simulation time. Gravity exceeds the thermal and magnetic counterparts at densities above $\rho\sim10^{-22}\,\mathrm{g\,cm}^{-3}$ ($n\sim40\,\percc$). The shaded area is bounded by the 25 and 75 percentile with a spread of about an order of magnitude at low densities and a factor of a few at high densities.
    }
    \label{fig:gradient-ratios}
\end{figure}
\begin{figure}
    \centering
    \includegraphics[width=8cm]{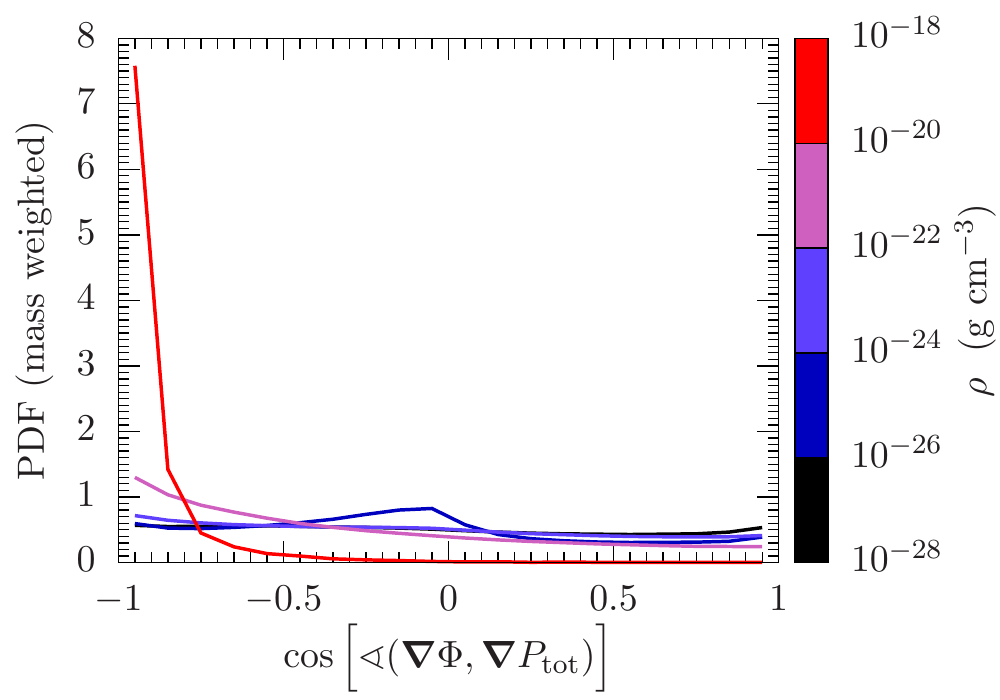}
    \caption{Distribution of the angle between the gravitational acceleration and the total pressure for run \texttt{B3-0.25pc} at the end of the simulation time. At low densities the angle is uniformly distributed. Above $\rho\sim10^{-20}\,\mathrm{g\,cm}^{-3}$ ($n\sim4000\,\percc$) the gravitational acceleration is predominantly anti-parallel to the total pressure force.}
    \label{fig:angle-acc-Ptot-PDF}
\end{figure}
\begin{figure}
    \centering
    \includegraphics[width=8cm]{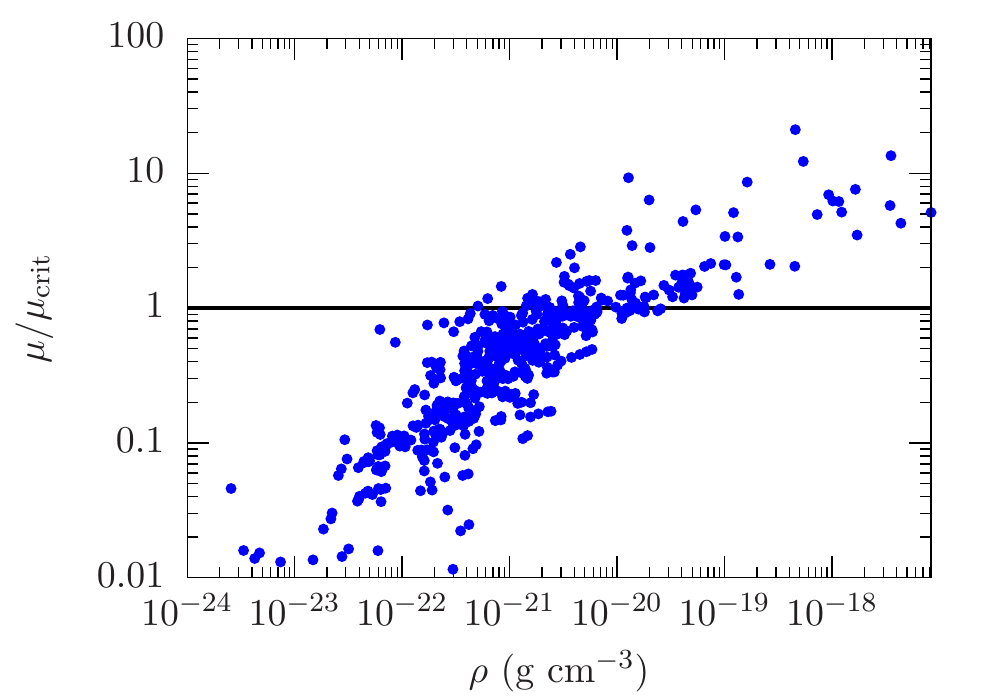}
    \caption{Mass-to-flux-ratio in units of the critical value as a function of density for the selected clouds in run \texttt{B3-0.25pc} at the end of the simulation time. Above a density of $\rho\sim10^{-21}-10^{-20}\,\mathrm{g\,cm}^{-3}$ ($n\sim400-4000\,\percc$) the clouds reach the critical mass-to-flux ratio for gravitational instability.}
    \label{fig:mass-to-flux-ratios}
\end{figure}

In order to further assess the connection between gravitational instability and change of orientation, we investigate the magnitude of the gradients of the individual pressure components that act as accelerators separately, i.e. the thermal, $\vnabla P_\mathrm{th}$, and magnetic, $\vnabla P_\mathrm{mag}$, pressure gradients as well as the gravitational force per unit volume, $\rho\vnabla\Phi$. We note that the reduction of the magnitude of the gradients is a simplification that cannot definitely prove gravitationally driven contraction because in principle the thermal and magnetic pressure gradients can locally assist gravity instead of opposing it. For the magnetic component we also neglect magnetic tension because it is of the same order of magnitude as the magnetic pressure, see appendix~\ref{sec:app-magnetic-tension}. We show the ratio of gravitational force per unit volume $\rho\vnabla\Phi$ over the individual pressure contributions in Fig.~\ref{fig:gradient-ratios} as a function of density, where $P_\mathrm{kin} = \rho\vektor{v}^2$. The lines present the median of the distribution along the ordinate with the shaded area being bounded by the 25 and 75 percentile. Above $\rho\sim10^{-22}\,\mathrm{g\,cm}^{-3}$ ($n\sim40\,\percc$) the gravitational force  dominates over the thermal and magnetic counterparts. At a density of $\rho\sim10^{-20}\,\mathrm{g\,cm}^{-3}$ ($n\sim4000\,\percc$) the force contribution due to gravity is already an order of magnitude stronger than the one by magnetic pressure. This suggests that magnetic forces are unable to dominate the geometry of the flow even if magnetic tension would be included in the analysis. The kinetic component dominates over the gravitational one except for the highest densities. The plot also indirectly indicates that the kinetic pressure gradient dominates over the thermal one for all densities and that it dominates over the magnetic counterpart above $\rho\sim10^{-23}\,\mathrm{g\,cm}^{-3}$ because the kinetic curve is below the two others. We note that the overlap within the shaded ranges of the pressure \emph{gradients} corresponds to the approximate equipartition in the \emph{energy densities}. As the transition in orientation with respect to the magnetic field occurs almost precisely at the density at which gravity dominates over magnetic and thermal acceleration, it is likely that gravitational contraction is the important driver that adjusts the magnetic field to the direction of the least resistance for the flow.

We also plot the distribution of angles between the gravitational attraction, $-\vnabla\phi$, and the total pressure gradient, $\vnabla P_\mathrm{tot}$, in Fig.~\ref{fig:angle-acc-Ptot-PDF}. In this case we would like to distinguish between a parallel and anti-parallel orientation, i.e. we do not reduce the distribution of $\cos[\sphericalangle(\vnabla\Phi, \vnabla P_\mathrm{tot})]$ to its modulus. The different lines indicate different density regimes. For all densities below $\rho\sim10^{-22}\,\mathrm{g\,cm}^{-3}$ ($n\sim40\,\percc$) the angles are approximately uniformly distributed. Above $\rho\sim10^{-20}\,\mathrm{g\,cm}^{-3}$ ($n\sim4000\,\percc$) a strong peak at negative unity indicates that the gravitational attraction is acting against the total pressure gradient.

The gravitational instability of dense clouds can be connected to the Jeans instability in the case of thermal pressure support, or with the transition from sub-critical to super-critical mass-to-flux ratio in the case of magnetic pressure support \citep[see e.g.][for recent reviews]{HennebelleInutsuka2019, GirichidisEtAl2020b}. For the identified clouds in simulation \texttt{B3-0.25pc} we plot the mass-to-flux ratio as a function of maximum cloud density in Fig.~\ref{fig:mass-to-flux-ratios} at the end of the simulation time. Despite a large scatter in the data there is a clear trend for low-density clouds to be supported by magnetic pressure. Clouds with central densities above $\rho\sim10^{-20}\,\mathrm{g\,cm}^{-3}$ ($n\sim4000\,\percc$) lie above the threshold value $\mu_\mathrm{crit}$ and cannot oppose gravitational collapse. We chose an analysis radius of $10\,\mathrm{pc}$, but variations from 2 to 15 pc do not change the result perceptibly. 


\section{Column density effects}
\label{sec:coldens-effects}

\begin{figure*}
    \begin{minipage}{\textwidth}
    \centering
    \includegraphics[width=\textwidth]{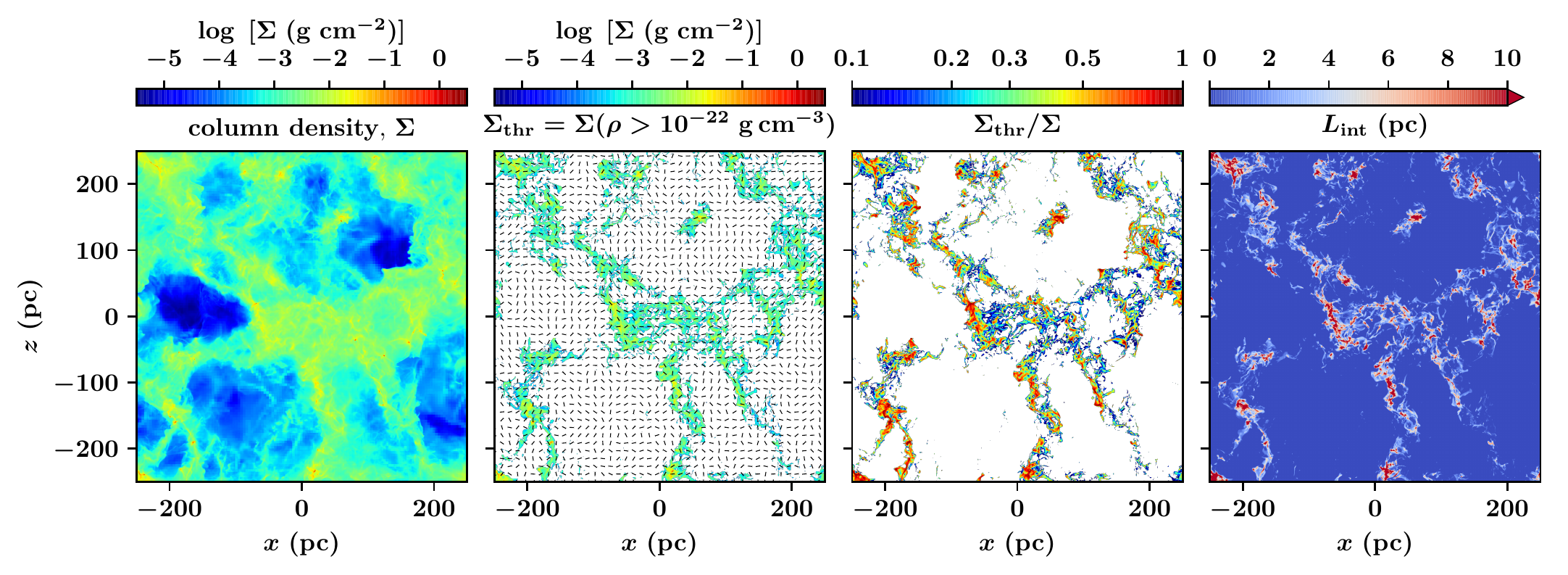}
    \caption{Column densities and integration length for the projection along $z$ for simulation \texttt{B3-0.25pc} at $t=20.8\,\mathrm{Myr}$. In the left panel we show the total column density (eq.~\eqref{eq:coldens}), in the second one the column density of high density gas (eq.~\eqref{eq:coldens-hi}), and in the third one the ratio of both. The right-hand panel depicts the integration length $L_\mathrm{int}$. In the second panel we also overplot the projected mass weighted magnetic field orientation as line segments.}
    \label{fig:coldens-comparison-ratio}
    \end{minipage}
\end{figure*}

\begin{figure}
    \centering
    \includegraphics[width=8cm]{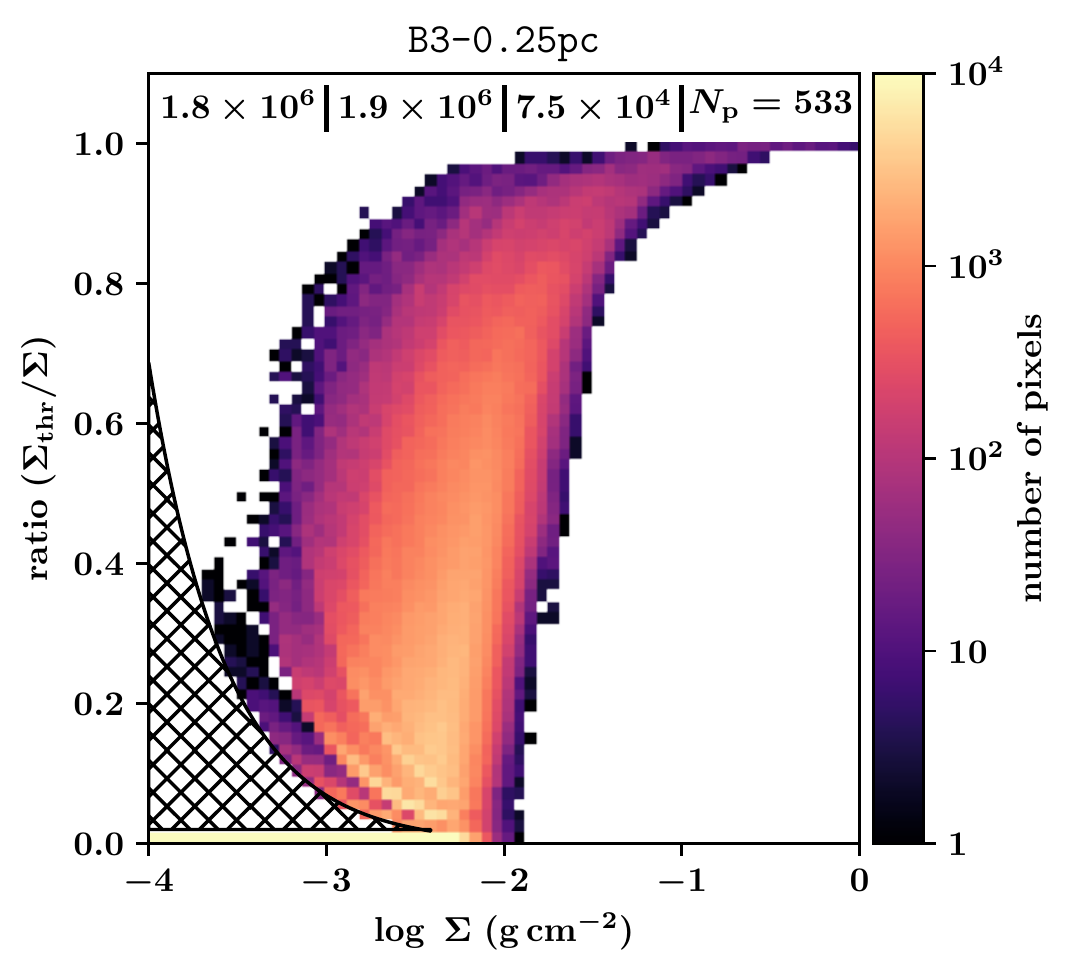}
    \caption{Fraction of the column density that originates from high density gas as a function of total column density for simulation \texttt{B3-0.25pc} at $t=20.8\,\mathrm{Myr}$. Colour coded is the number of pixels, with the numbers also shown at the top. High density gas above $\rho=10^{-22}\,\mathrm{g\,cm}^{-3}$ ($n\sim40\,\percc$) does not contribute to the lowest values column densities. For $10^{-3}\lesssim\Sigma/\mathrm{g\,cm}^{-2}\lesssim0.03$ ($4\times10^{20}\lesssim N/\mathrm{cm}^{-2}\lesssim\times10^{22}$) the high density gas only partially contributes to $\Sigma$. The high-$\Sigma$ end of the distribution is solely correlated with gas above $\rho=10^{-22}\,\mathrm{g\,cm}^{-3}$ ($n\sim40\,\percc$). The hatched region indicates impossible values (see text).}
    \label{fig:fraction-coldens}
\end{figure}

\begin{figure*}
    \begin{minipage}{\textwidth}
    \centering
    \includegraphics[width=\textwidth]{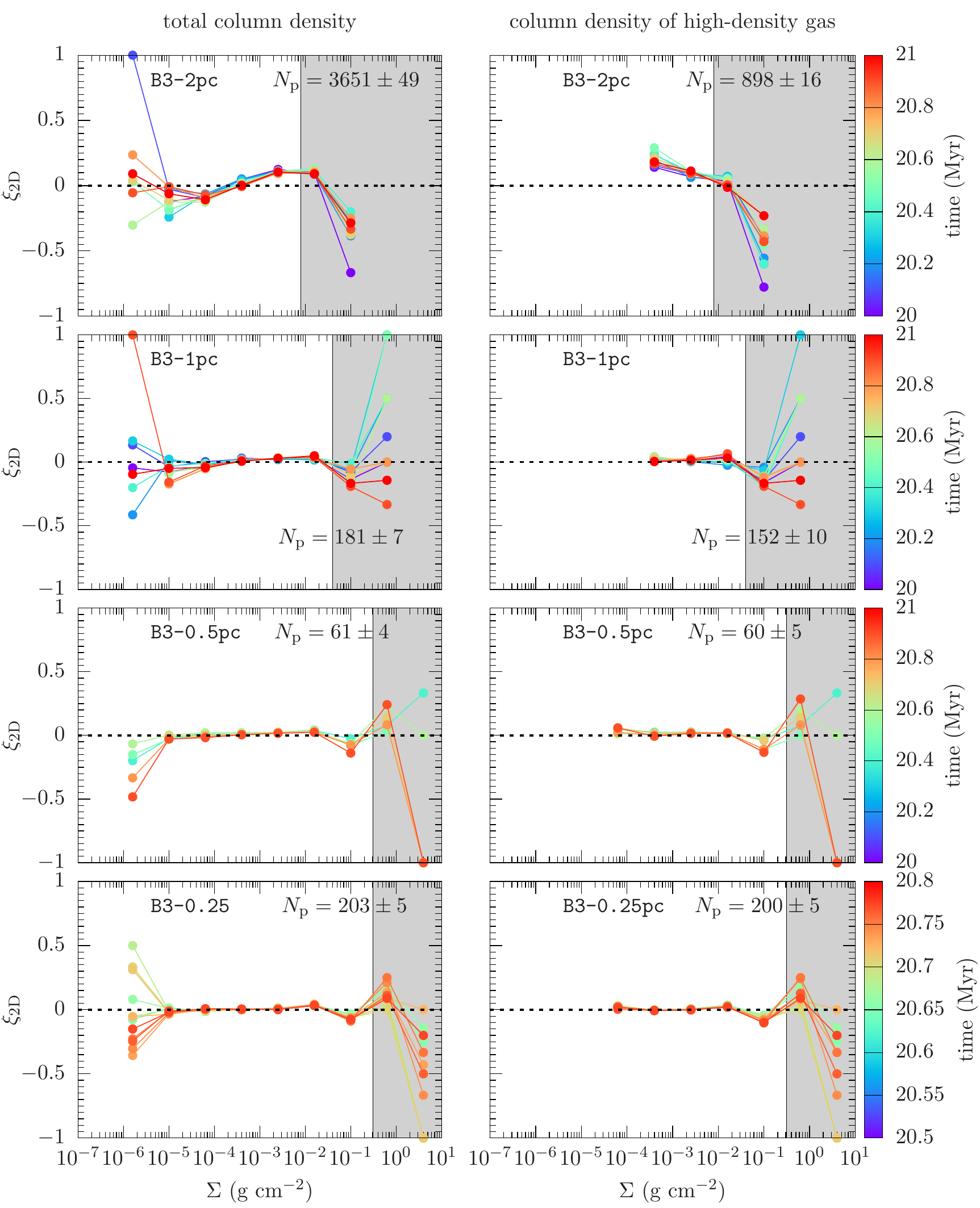}
\caption{Two-dimensional orientation parameter $\xi_\mathrm{2D}$ as a function of column density for $\Sigma$ (left-hand side) and $\Sigma_\mathrm{thr}$ (right-hand side). From top to bottom we increase the resolution. Colour coded are the different time snapshots. Note the different temporal range at the highest resolution simulation. We highlight the number of column density pixels, $N_\mathrm{p}$, that are taken into account to determine the two highest bins (shaded area). The temporal variation is indicated as the standard deviation.}
    \label{fig:xi-2D-resolution}
    \end{minipage}
\end{figure*}

\begin{figure}
    \centering
    \includegraphics[width=8cm]{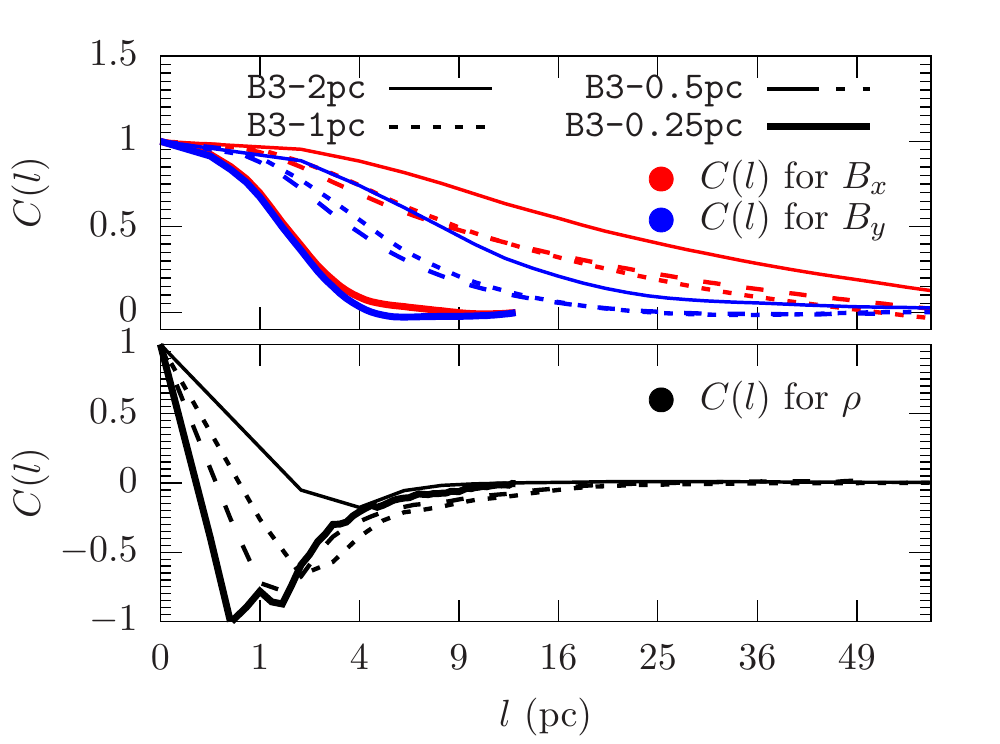}
    \caption{Correlation functions for the magnetic field and the density. The black lines indicate the density, blue curves represent $B_y$ and red curves $B_x$. Increasing the resolution reduces the correlation length.}
    \label{fig:correlation-functions}
\end{figure}

\begin{figure*}
    \begin{minipage}{\textwidth}
    \centering
    \includegraphics[width=\textwidth]{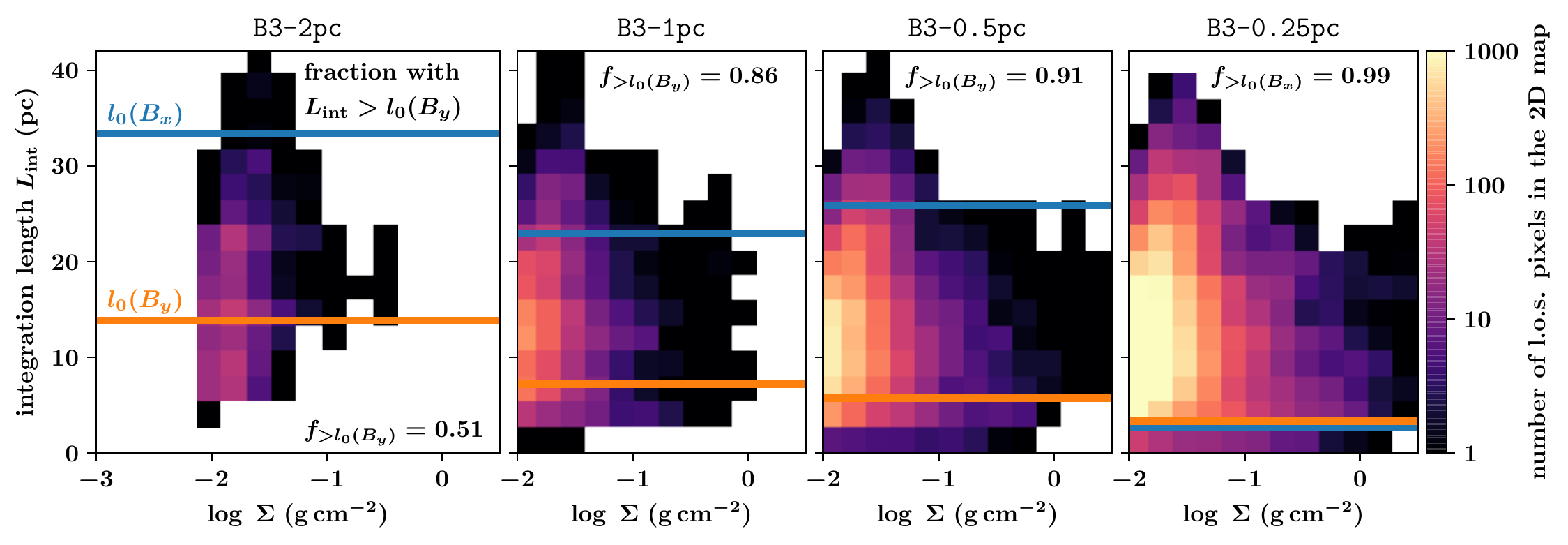}
    \caption{Two-dimensional histograms as a function of column density and integration length. Colour coded is the number of line-of-sight pixels in the 2D integrated maps. From left to right we show the time averaged data for the different resolutions. The correlation length of the magnetic field is indicated with the horizontal lines. We also indicate the fraction of projected line-of-sight elements for which the integration length exceeds the smallest correlation length. For all but the highest resolution $l_0(B_y)<l_0(B_x)$.}
    \label{fig:integration-length-res-comp}
    \end{minipage}
\end{figure*}

\subsection{Projected quantities}

Observations measure the angle between the magnetic field vector in the \emph{plane of the sky} and the \emph{column density}. Being a simple integrated quantity, the column density is straightforward to extract from simulations. Contrary, the magnetic field is not additive, so the field in the plane of the sky is ambiguous, i.e. it is not obvious which region along the line of sight the observed field originates from and how many cancellations due to field reversals occur. Intuitively, the mass-weighted average field along the line of sight seems to be a natural choice to appropriately reflect the correlation with the column density. The correlation between the observed magnetic field and the corresponding gas structures is complex and beyond the scope of this paper. A more detailed analysis in terms of the radiation processes like the amount of actually observable gas is discussed in \citet{SeifriedEtAl2019b}. Details of polarization in clouds are presented in \citet{SeifriedEtAl2019a}.

In order to investigate the projection effects we compute the column density
\begin{equation}
\label{eq:coldens}
    \Sigma = \sum_{i}\,\rho_i\Delta z_i,
\end{equation}
as well as the column density of gas above a threshold density, $\rho_\mathrm{thr}$,
\begin{equation}
\label{eq:coldens-hi}
    \Sigma_{\mathrm{thr}} = \sum_{i}\,\theta(\rho_i-\rho_{\mathrm{thr}})\,\rho_i\Delta z_i.
\end{equation}
Here, the cell index $i$ includes all cells along the line of sight, $\Delta z_i$ and $\rho_i$ are the size of a cell along the line of sight and the corresponding density, and $\theta$ is the Heaviside step function. We also define the integration length corresponding to $\Sigma_{\mathrm{thr}}$ as
\begin{equation}
	\label{eq:int-length}
    L_\mathrm{int} = \sum_{i}\,\theta(\rho_i-\rho_\mathrm{thr})\,\Delta z_i.
\end{equation}
We restrict our analysis to the projection along $z$, i.e. a face-on view of the mid-plane. This direction reduces the effects of integrating along several individual clouds and therefore marks the least complicated correlation between three-dimensional and projected quantities. It is also the direction that probes the entire domain in a useful manner. By separately investigating the column density of the dense gas we ensure that mainly gas with a noticeable alignment ($\vnabla\rho\parallel\vektor{B}$) is considered. We set the threshold value to $\rho_\mathrm{thr}=10^{-22}\,\mathrm{g\,cm}^{-3}$ ($n\sim40\,\percc$). For the magnetic field -- here reduced to a two-dimensional vector in the $xy$-plane -- we use the density weighted field strength along the line of sight
\begin{equation}
    \label{eq:mag-field-mw}
    \vektor{B} = \Sigma^{-1}\,\sum_{i} \vektor{B}_{i}\,\rho_i\,\Delta z_i
\end{equation}
and the analogous quantity to $\Sigma_\mathrm{thr}$,
\begin{equation}
    \label{eq:mag-field-hi}
    \vektor{B}_{\mathrm{thr}} = \Sigma_{\mathrm{thr}}^{-1}\,\sum_{i} \vektor{B}_i\,\theta(\rho_i-\rho_{\mathrm{thr}})\,\rho_i\Delta z_i,
\end{equation}
where only the magnetic field in the high-density gas is considered for the density weighted average. Fig.~\ref{fig:coldens-comparison-ratio} shows the column density ($\Sigma$, left-hand panel), the high-density counterpart ($\Sigma_\mathrm{thr}$, second panel), the ratio $\Sigma_\mathrm{thr}/\Sigma$ (third panel) as well as the integration length for $\Sigma_\mathrm{thr}$ for every line of sight along $z$ (right-hand panel) for simulation \texttt{B3-0.25pc} at $t=20.8\,\mathrm{Myr}$ using the threshold density of $\rho_\mathrm{thr}=10^{-22}\,\mathrm{g\,cm}^{-3}$ ($n\sim40\,\percc$). In the second panel we indicate the orientation of the mass weighted projected magnetic field in the plane of the sky as line segments. The direct integration ($\Sigma$) shows the largest values and a column density range of six orders of magnitude. In $\Sigma_\mathrm{thr}$ the high-density gas mainly contributes to intermediate and high column densities. There is no obvious visible correlation between the projected magnetic field and the column density stemming from high density gas. However, the ratio in the right-hand panel spans a dynamic range from $0.1-1$, which illustrates that in some regions the high-density gas only makes a small share of the total column. For $L_\mathrm{int}$ we find connected patches with large integration lengths ($>10\,\mathrm{pc}$) and a filamentary network of regions with $L_\mathrm{int}$ of a few parsec.

A more quantitative distribution of the column density is plotted in Fig.~\ref{fig:fraction-coldens}, which shows the fraction that the high-density gas ($\rho>10^{-22}\,\mathrm{g\,cm}^{-3}$) contributes to the total column density as a function of $\Sigma$. Colour coded is the number of cells in the column density map. In addition we show the total number of pixels within one decade of column density, i.e. the integrated number of the colour coded distribution. Low column densities are dominated by low-density gas. For the intermediate regime the contribution from high-density gas spans the entire range. The very high column densities are dominated by high-$\rho$ contributions. The hatched region indicates the regime of impossible values. The limit of this region is given by the minimum column density that originates from a single cell at the threshold density, $\Sigma_\mathrm{thr, min}=\rho_\mathrm{thr}\Delta z_\mathrm{min}=7.7\times10^{-5}\,\gperscm$ ($N_\mathrm{thr}\sim10^{19}\,\perscm$), where $\Delta z$ is the minimum cell size. The distribution is therefore bounded by the function $\Sigma_\mathrm{thr, min}/\Sigma$. Below that curve $\Sigma_\mathrm{thr}=0$.

\subsection{$\xi_\mathrm{2D}$}

The scaling of $\xi_\mathrm{2D}$ as a function of column density is shown in Fig.~\ref{fig:xi-2D-resolution} for the total column density (left-hand side) and the column density of the dense gas (right-hand side) for different resolutions from $2\,\mathrm{pc}$ down to $0.25\,\mathrm{pc}$ (top to bottom, see definition of $\xi_\mathrm{2D}$ in Eq.~\ref{eq:xi_2D}). The different lines indicate the time evolution (colour coded). The shaded area shows the two highest column density bins with the corresponding number of column density pixels $N_\mathrm{p}$ used to derive $\xi_\mathrm{2D}$. For simulation \texttt{B3-2pc} this range covers the column densities above $\Sigma\sim10^{-2}\,\gperscm$ ($N\ge4\times10^{21}\,\perscm$). The two highest column density bins for \texttt{B3-1pc} are above $\Sigma\sim0.03\,\gperscm$ ($N\sim10^{22}\,\perscm$) and for the two runs with the highest resolution $N_p$ includes regions with $\Sigma\sim0.3\,\gperscm$ ($N\sim10^{23}\,\perscm$). The number shows the mean over all time snapshots with the deviation being the standard deviation over time. We note that in the low-resolution run \texttt{B3-2pc} the high column densities ($\Sigma\gtrsim10^{-2}\,\mathrm{g\,cm}^{-2}$, $N\sim4\times10^{21}\,\perscm$) show negative values for $\xi_\mathrm{2D}$ for all times reflecting the alignment that we measured in the 3D quantities. Increasing the resolution by a factor of two yields both positive and negative $\xi_\mathrm{2D}$ changing within a time scale of $1\,\mathrm{Myr}$ in a more fluctuating than systematic manner. A resolution of $0.5\,\mathrm{pc}$ indicates an equally fluctuating behaviour. The differences between $\Sigma$ and $\Sigma_\mathrm{tr}$ are minor. Given the strongly varying results for large column densities we are not able to connect the clear signal of the orientation in the 3D quantities with the projected values and the magnetic field in the plane of the sky. The bottom plots show the time evolution for simulation \texttt{B3-0.25pc}. The values for the highest column densities also vary over time with non-systematic fluctuations. Here, all values at the penultimate column density are positive, while all numbers at the highest column density are negative. At this point we would like to stress that the total number of column density data points at the highest bins is relatively small ranging from 10 to 25 for each snapshot. As a consequence $\xi_\mathrm{2D}$ at the highest column density bin can easily fluctuate due to the low number statistics and we cannot draw a solid conclusion from this number alone. However, even when ignoring the highest column density bin, the penultimate one and the third to last (i.e. the transition from the white to the grey shaded area) do not indicate a stable signal over time. Here, the number of data points used to derive $\xi_\mathrm{2D}$ are the lowest for \texttt{B3-0.5pc} (60-600 data points) and increase to ranges from 150-$10^4$ data points, which should be sufficient. Instead of simply relying on the number of data points and how probable the results for $\xi_\mathrm{2D}$ at high column densities are we would like to investigate the local conditions further to understand the effects along the line of sight.

To put the temporal fluctuations and the corresponding analysis time into perspective, we estimate how strong the dynamical changes in the gas and magnetic field structure could be during this time span. If this analysis time span is longer than a turbulent crossing time, the dense structures can be reshaped significantly and/or rotated, such that the projected appearance differs. At median velocity dispersions of $3\,\mathrm{km\,s}^{-1}$ in the dense gas above $\rho_\mathrm{thr}$ the turbulent crossing time for $1\,\mathrm{pc}$ is approximately $0.3\,\mathrm{Myr}$, so the analysis time covers approximately three crossing times in the case of simulation \texttt{B3-1pc}. At the densest regions the gas structures and the projected signal are thus not unlikely to change because of turbulent dynamics and gravitational collapse. At the highest resolution the crossing time is $82\,\mathrm{kyr}$, which is comparable to the analysis time. In this run, the regions at the highest density and highest resolution are therefore less susceptible to strong changes in the gas structures. The fact that all numbers at the highest column densities are negative could thus be due to the short analysis time and reflect a short snapshot in time. Over a longer evolution and more dynamical changes there is the possibility for positive values if $\xi_\mathrm{2D}$ depends sensitively on the specific line of sight. During the analysis time the values change from negative unity to zero in a non-monotonic manner. We note that in 3D the dynamical evolution is directly connected to the forces and pressures. The signal in 3D can thus be stable over time even though the projected effect is not.

\subsection{Correlation lengths}

In order to get further insights into $\xi_\mathrm{2D}$ we return to the three-dimensional data and investigate the changes in the signal along the line of sight that leads to $\xi_\mathrm{2D}$. Of particular importance are the changes of the magnetic field orientation. We analyse two different quantities, namely the correlation length as a known turbulence measure and the distribution of angles along the line of sight.

First, we compare the integration length along the line of sight with the correlation length of the gas and magnetic field structures. If the integration length that leads to a certain $\Sigma$ is comparable to or larger than the correlation length of the density or the magnetic field, the signal could simply be averaged out or alters stochastically, which can explain the fact that $\xi_\mathrm{2D}$ shows positive and negative values. We use the integration length for the high-density, see Eq.~\ref{eq:int-length}. We investigate the correlation length of the field in the dense regions ($\rho>10^{-22}\,\mathrm{g\,cm}^{-3}$) using the second order structure function \citep[e.g.][]{MoninIaglom1975, SheLeveque1994}
\begin{equation}
    D(l) = D(|\vektor{l}|) =
    {\langle \left[f(\vektor{x}+\vektor{l})-f(\vektor{x})\right]^2\rangle}_{\vektor{l}},
\end{equation}
with the three-dimensional position $\vektor{x}$ and distance $\vektor{l}$, which we bin radially to obtain the isotropic structure function, $D(l)$. The averaging is performed over position $\vektor{x}$ and over lag vectors $\vektor{l}$ of a given length. We compute the normalized autocorrelation function
\begin{equation}
    C(l) = 1-\frac{D(l)}{2\sigma^2},
\end{equation}
where the normalization $2\sigma^2$ is the value of $D(l)$ at which the function $f(\vektor{x})$ is no longer correlated and $\sigma$ is the dispersion of $f$. Finally the correlation length is given by \citep[e.g.][]{MatthaeusEtAl1995,Shalchi2008,HollinsEtAl2017}
\begin{equation}
\label{eq:correlation-length}
    l_0 = \int_0^{\infty} C(l)\,\mathrm{d} l. 
\end{equation}
The value of $l_0$ depends on the upper integration boundary as well as on the shape of the integrand and thus on $\sigma$. In turbulence simulations the boxes are typically evolved for many turnover times in fully periodic boxes. The correlation function is also computed at every position in the simulation box (or a representative sample of it) independent of the density. Using a periodic box and integrating up to the box length hence gives a natural choice for the upper limit of the integration boundary. In our case the upper limit needs to be set to a reasonable value smaller than the box length in order to find the correlation length of the \emph{dense} gas. This means we can not strictly fulfil the upper limit in the definition at $\infty$ and also the convergence of $C(l)$ at our upper integration limit needs to be taken with caution. We therefore emphasise that we refrain from taking the value of $l_0$ as a precise measure but rather as an estimate at which the data (density, magnetic fields) become uncorrelated (see also discussion in \citealt{GentEtAl2013b} and \citealt{HollinsEtAl2017}). We set the upper integration limit to $60\,\mathrm{pc}$ for simulations \texttt{B3-2pc}, \texttt{B3-1pc}, \texttt{B3-0.5pc}, which proved to be a reasonable choice after some tests with larger and smaller radii. For simulation \texttt{B3-0.25pc} the typical length scales are smaller and a shorter integration length of $\sim15\,\mathrm{pc}$ has proven to give the most stable results. To measure $D(l)$ we chose all gas cells with densities above the threshold density and investigate a region around that cell within the integration radius. We show the correlation function for the density and the two relevant magnetic field components ($B_x$, $B_y$) in Fig.~\ref{fig:correlation-functions} for all simulations at $t=20.8\,\mathrm{Myr}$. The density indicates a very short correlation length with significant fluctuations within a radius of a few computational cells (black lines). The magnetic field along $x$ shows the largest correlation (red lines). This reflects the initial conditions, because we initially align the field in $x$-direction. The $y$-component of the field is initially set to zero. The magnetic structures for $B_y$ (blue lines) are thus solely a result of the turbulent driving of SNe and gravitational interaction. All $C(l)$ asymptote to zero within the integration range except for the case of $B_x$ in simulation \texttt{B3-2pc}. In this case $C(l)$ is not converged up the radius of $60\,\mathrm{pc}$. Since the $x$ component is still dominated by the initially uniform field that spans the entire domain, we simply note that the correlation length is larger than typical sizes of density structures like molecular clouds and do not further investigate the convergence for this simulation. Overall, we note that $C(l)$ for the density converges at least as fast as the curves for the magnetic field. This highlights that the local gas structures vary on even shorter spatial scales. With increasing resolution we find correlation functions that are stronger peaked indicating a smaller correlation length. The integrals for $l_0$ yield $l_0(B_x) = 33$, $23$, $26$ and $5\,\mathrm{pc}$ for simulations \texttt{B3-2pc}, \texttt{B3-1pc}, \texttt{B3-0.5pc} and \texttt{B3-0.25pc}, respectively. The corresponding values for $B_y$ read $l_0(B_y) = 14$, $7$, $6$ and $3\,\mathrm{pc}$. For the highest resolution the correlation lengths for $B_x$ and $B_y$ are remarkably similar. This indicates that the local magnetic structures around the dense cores are decoupled from the initial large scale field along the x-direction.

Alternatively, one can assume the autocorrelation function to have a Gaussian shape and fit for the correlation length $l_0$
\begin{equation}
	C'(l) = \exp\left(-\frac{l^2}{l_0^2}\right).
\end{equation}
The corresponding values for the fits read $l_0(B_x) = 29$, $14$, $15$ and $2\,\mathrm{pc}$ for simulations \texttt{B3-2pc}, \texttt{B3-1pc}, \texttt{B3-0.5pc} and \texttt{B3-0.25pc}, respectively, and $l_0(B_y) = 11$, $6$, $5$ and $2\,\mathrm{pc}$. We emphasize that the obtained correlation lengths can differ by a factor of approximately two. None the less, we find a systematic decrease in correlation length with increasing resolution. This is intuitively in agreement with the paradigm of smaller turbulent eddies that cause field entanglement on scales of the turbulence combined with the gravitational collapse on scales down to the resolution limit.

Figure~\ref{fig:integration-length-res-comp} summarises the connection between $\Sigma_{\mathrm{thr}}$, $L_\mathrm{int}$ and $l_0$. For this comparison we stick to the overall larger values of $l_0$ obtained from the first method, because this will be the more conservative estimate. The four panels from left to right show the data for simulations \texttt{B3-2pc}, \texttt{B3-1pc}, \texttt{B3-0.5pc} and \texttt{B3-0.25pc}. Plotted is the integration length as a function of the column density. Colour coded is the number of line of sight pixels in the two-dimensional map. We would like to emphasize that the integration length is computed based on the dense gas, which is predominantly in cells at the highest refinement level. For an integration length of $L_\mathrm{int}=10\,\mathrm{pc}$ each line of sight passes approximately 5, 10, 20, and 40 cells for the runs \texttt{B3-2pc}, \texttt{B3-1pc}, \texttt{B3-0.5pc} and \texttt{B3-0.25pc}, respectively. The horizontal lines indicate the correlation length of $B_x$ and $B_y$ as defined in Eq.~\eqref{eq:correlation-length}. Focusing on the high-column density regime ($\Sigma\ge10^{-2}\,\mathrm{g\,cm}^{-2}$, $N\ge4\times10^{21}\,\perscm$) we note that for most of the projected pixels, the integration length is smaller than the correlation length of $B_x$. An exception is the highest resolution simulation, which has a very small correlation length of $B_x$. More importantly, for almost all pixels, $L_\mathrm{int}$ is larger than the correlation length of $B_y$. This means that in almost all cases the orientation of the $y$-component changes several times along the line of sight and is thus averaged out along the integration path. Consequently, the correlation between the gas structures and the field can easily be dominated by fluctuations and appear as a random signal. The reason for why in the low resolution case the value of $\xi_\mathrm{2D}$ is more stable over the investigated time is due to the fact that the total number of cells over which we integrate along the line of sight is relatively small. The maximum number of cells is $\lesssim30$ for column densities above $\sim0.03\,\mathrm{g\,cm}^{-2}$. This means that the contributions to the total column density measurement stem from only a few cells, so does the measurement of the magnetic orientation, which explains the remaining correlation. In the runs with higher resolution $\xi_\mathrm{2D}$ includes a total number of cells $\gtrsim100$ and the 3D signal of the alignment can be weakened or lost. We note however, that the total number of pixels in the colour coded map is \emph{not} along one individual line of sight.

\subsection{Centroid method}

\begin{figure}
\centering
\includegraphics[width=8cm]{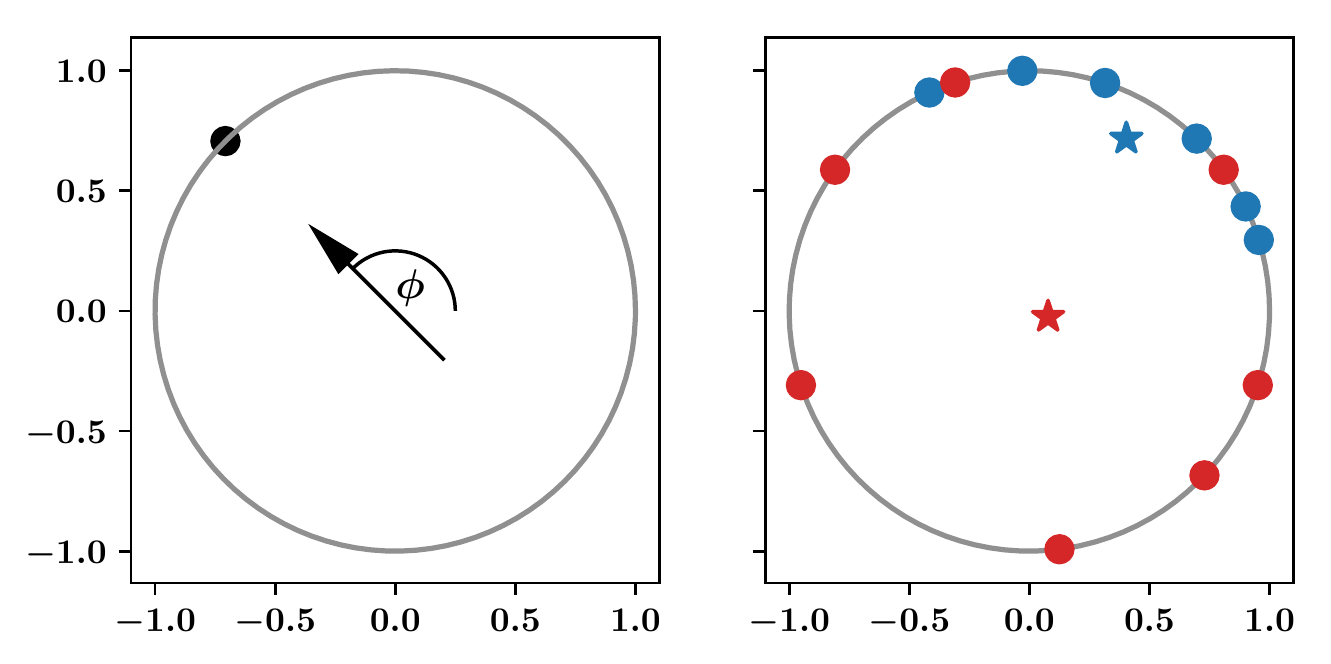}
\caption{Illustration of the centroid method to quantify the dispersion of the angles. For each cell we assign the angle $\phi$ of the magnetic field vector in the plane of the sky to its corresponding point on the unit circle (left-hand panel). For each line of sight we compute the centroid of all angles (right-hand panel). For a narrow distribution of angles the centroid is located further away from the centre (blue example) compared to a large spread of angles (red example).}
\label{fig:centroid-sketch}
\end{figure}

\begin{figure}
\centering
\includegraphics[width=7.7cm]{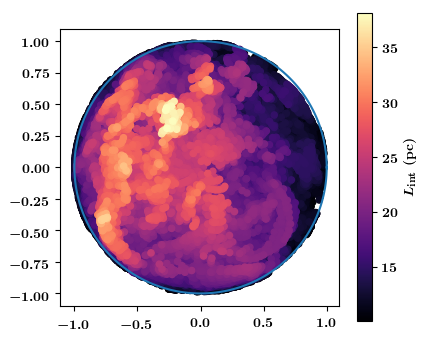}\\
\includegraphics[width=7.7cm]{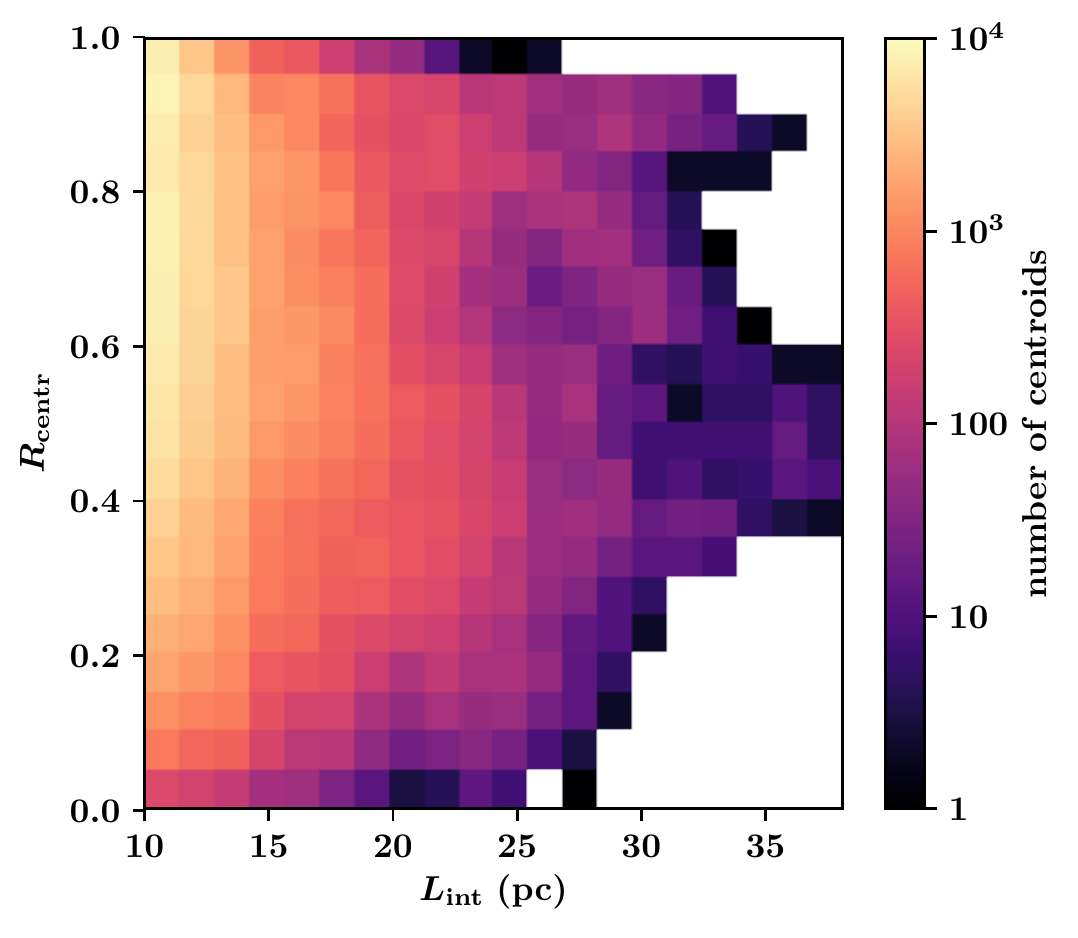}\\
\includegraphics[width=7.7cm]{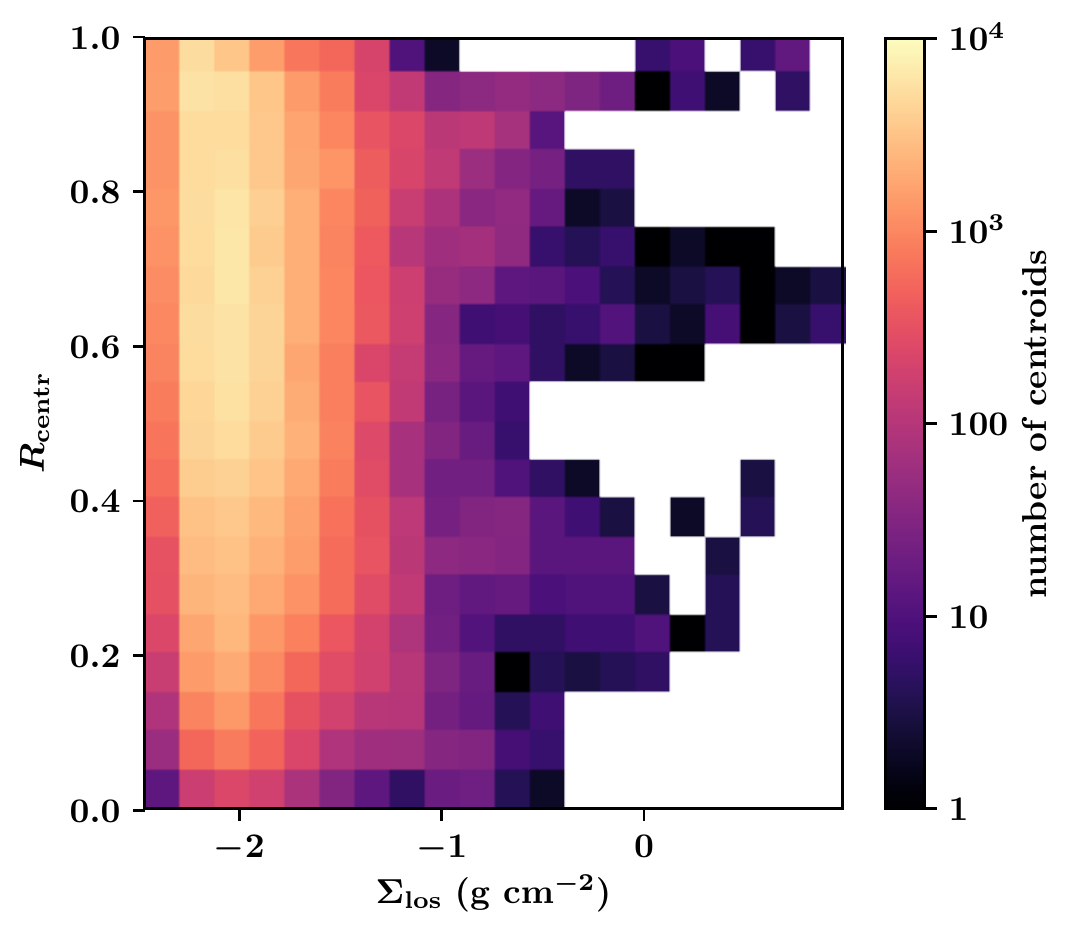}\\
\caption{Dispersion of the orientation of the magnetic field using the centroid method. Shown are data for simulation \texttt{B3-0.25pc} analysed over the entire analysis time. In the top panel each point corresponds to the centroid of one line of sight with the colour representing $L_\mathrm{int}>20\,\mathrm{pc}$. The middle and bottom panels show two-dimensional histograms of the number of line of sight rays as a function of radius of the centroid $R_\mathrm{centr}$ versus $L_\mathrm{int}$ (middle) and $\Sigma_\mathrm{los}$ (bottom). We find a broad distribution indicating regions of large dispersion of the angles (centroid close to centre; small $R_\mathrm{centr}$) as well as centroids close to the unit circle with radii close to unity. For both quantities ($L_\mathrm{int}$ and $\Sigma_\mathrm{los}$) we find a broad distribution covering the full range of the centroid radii.}
\label{fig:centroid-method}
\end{figure}

As a second analysis we directly measure the dispersion of the magnetic field orientation along the line of sight, specifically for $\Sigma_\mathrm{thr}$ and the corresponding integration length $L_\mathrm{int}$ and column density, respectively. Since the distribution of angles is periodic, we cannot simply compute a width of the distribution like the standard deviation. Instead, we use the centroid method as outlined in \citet{ForbesAlonso2001}. We assign the orientation of the magnetic field vector with angle $\phi$ in the plane of the sky to a point on the unit circle as illustrated in the left-hand panel of Fig~\ref{fig:centroid-sketch}. We do this for all cells with $\rho>\rho_\mathrm{thr}$ along an individual line of sight. For this line of sight we then compute the centroid of the angular distribution as illustrated in the right-hand panel of Fig.~\ref{fig:centroid-sketch}. If the angle does not change much along the line of sight, the positions on the unit circle are located in a small range and the resulting centroid is close to the circle, i.e. at a radius close to unity. This configuration is illustrated with the blue points and the blue star as the centroid. Contrary, if the angles vary a lot (red symbols), the centroid is close to the centre of the unit circle. We highlight that both for the dynamical effects in ideal MHD as well as for the simple relative orientation between the field and the gas gradients we do not worry about the actual pointing direction of the magnetic field vector, i.e. we do not need to distinguish between $\vektor{B}$ and $-\vektor{B}$. Consequently, we can restrict our analysis to a range from 0 to $\pi$, such that if the field rotates by $\pi$ we cannot distinguish any more between the original and the rotated field. In order to compute a centroid and visualize the spread in angles we stretch the range to 0 to $2\pi$, i.e. every angle $\phi$ is multiplied by two.

We discuss the results of this method for simulation \texttt{B3-0.25pc} using the data of the entire analysis time. A comparison with different resolutions is shown in Appendix~\ref{sec:centroid-resolution}. In Fig.~\ref{fig:centroid-method} we show the unit circle with the centroid points for each line of sight with $L_\mathrm{int}\ge10\,\mathrm{pc}$ in the top panel. The integration length is colour coded. The middle panel depicts the two-dimensional histogram of the number of line-of-sight rays with a measured $L_\mathrm{int}$ and centroid radius $R_\mathrm{centr}$. The bottom panel shows the corresponding distribution as a function of $\Sigma_\mathrm{los}$. We note that all regions are populated both with small $R_\mathrm{centr}$ and consequently a large angular dispersion as well as regions with centroids close to $R_\mathrm{centr}\approx1$, representing a small angular spread along the line of sight. Besides small local overdensities in the distributions there is no clear trend of the abscissa with $R_\mathrm{centr}$. In particular for the case of the column density (bottom panel) the almost uniformly distributed spread across $R_\mathrm{centr}$ highlights that the field can undergo significant rotation (or even many reversals) along each line of sight. This underlines the possible difficulty of finding a correlated orientation of the line of sight integrated field with the column density structures. This part of the analysis is consistent with the simple analysis of $\xi_\mathrm{2D}$ and the short correlation lengths of the field and the gas structures compared to the integration length.

Even if at every local 3D position the orientation of $\vektor{B}$ with respect to $\nabla\rho$ follows an idealised distribution similar to the one measured in $\xi_\mathrm{3D}$, the combination of a strongly varying angle and an integrated column density gradient is likely to give a somewhat random measurement for $\xi_\mathrm{2D}$. This emphasises that the correlation between the column density and the magnetic alignment might be hard to find in individual clouds.

\section{Discussion}
\label{sec:discussion}

\subsection{Galactic environment}
Our simulations focus on a patch of the ISM in a box with a length of $0.5\,\mathrm{kpc}$ and do not include the full galactic (magnetic) environment. We initially set the magnetic field along the $x$ axis, which mimics the large scale field following a spiral arm. During the evolution of the simulations the field is not connected to an external environment. A large scale, coherent galactic magnetic field in a larger simulation box or an entire galactic disc could stabilize the field, in particular in the low-density gas. This in turn could reduce the strong entanglement of the field in these regions and change the degree of alignment of the field with the density gradient or the velocity. A field that is even more aligned with the large scale structure of filaments and clouds (more peaked distribution towards a perpendicular orientation of the field with the density gradient) may therefore be possible. The high density regions -- in particular the gravitationally collapsing regions -- are dynamically decoupled from the flow and the field on scales of the box size. The main results of the paper are thus unlikely to be influenced by a missing large scale magnetic field of a few $\mu\mathrm{G}$.

\subsection{Comparison to other studies}

A number of idealised set-ups have been performed, which are in agreement with the results in our more self-consistent simulation. In a theoretical study, \citet{SolerHennebelle2017} derived an evolution equation of the angle $\phi$. By numerically estimating the terms in the evolution equation they concluded that aligned and perpendicular configurations are a natural consequence of the continuity equation and Faraday's law.

The transition density between the alignment regimes agrees very well in our simulations and observations \citep{AlinaEtAl2019, FisselEtAl2019}. In addition, the tangled structure of the magnetic field in dense turbulent regions as investigated in our simulations and the resulting difficulty in identifying the alignment in projection could be a plausible link to the missing detection in high-density filaments reported by \citet{AlinaEtAl2019}.

\citet{ChenKingLi2016} performed converging flow simulations that highlight the importance of gravitational collapse for the orientation of the magnetic field lines and the gas structure. Their simulation set-up focuses on smaller scales with a cubic box of $1\,\mathrm{pc}$ side length and a resolution down to $\Delta x=0.004\,\mathrm{pc}$. Their average initial density is at $10\,\percc$ ($\sim2\times10^{-21}\,\gpercc$) using an isothermal equation of state at a temperature of $10\,\mathrm{K}$. Using colliding flow simulations they can better control the average magnetization of the gas and the orientation of the initial magnetic field relative to the flow of the gas. However, the turbulent motions are not self-consistently generated together with the formation of dense and cold regions. Concerning the analysis \citet{ChenKingLi2016} discuss gravitational collapse as an important agent in changing the orientation of the field with respect to the gas structures. However, they do not analyse the individual forces directly.

A more detailed comparison between simulations and observations has recently been performed by \citet{SeifriedEtAl2020} using colliding flow simulations as well as a similar set of simulations as in this study. They report a flip from parallel to perpendicular orientation at high densities of $n\sim10^2-10^3\,\percc$ $\rho\sim2\times10^{-22}-2\times10^{-21}\,\gpercc$). It is suggested that this transition occurs at a mass-to-flux-ratio of order unit in agreement with our findings. For the 2D analysis they use radiative transfer calculations and Projected Rayleigh Statistics. A transition flip in the orientation is not always measured despite the more sophisticated analysis compared to this study. For the colliding flow simulations \citet{SeifriedEtAl2020} use a large range of initial magnetisation with a mass to flux ratio ranging from $0.54-4.3$ in units of the critical one. The change in orientation depends on the magnetisation where weak magnetic fields tend to have a parallel orientation (positive $Z_X$ in their notation). The change to negative $Z_X$ and a perpendicular orientation is only measured for strong magnetic fields. In their models the correlation between the 3D and 2D measurements is more stable and correlates better with one another. We speculate that this better correlation is likely caused by their initially uniform magnetic field that is mainly stirred by a one-time collision of the two flows. The resulting entanglement of the magnetic field and the formation of structures are directly connected with relatively short turbulent mixing times. This speculation is in agreement with their analysis of the SILCC-zoom models, which are very similar to our set-ups. \citet{SeifriedEtAl2020} report a weak trend of decreasing $Z_X$ (corresponding to our $\xi_\mathrm{2D}$) with the column density $\Sigma$. In a number of cases $Z_X$ does not reach negative numbers or only reaches negative numbers for a small range of column densities. The strongly differing results between different lines of sight and different molecular clouds is nicely illustrated in their Fig~7. The more complicated formation history of the entanglement of the magnetic field and the condensation of cold and dense gas out of the warm turbulent gas might still locally preserve the orientation of magnetic field and gas density gradients. The integrated 2D effect, however, can be cancelled out completely or depend on individual regions without an overall universal scaling over the investigated column density range. Overall, their results and their quoted uncertainty are well in line with our results of a short correlation length and a possibly large angular dispersion along the line of sight.

\citet{WuEtAl2020} perform simulations of two giant molecular clouds with a radius of $20\,\mathrm{pc}$ each. The clouds with an initial number density of $10\,\percc$ consist of cold gas ($t=10\,\mathrm{K}$). The comparably strong initial magnetic fields ($10-50\,\mu\mathrm{G}$) are uniform. They distinguish between non-colliding and a colliding set-up and investigate the magnetic field orientation in projection together with the column density structures. The combinations of their models reveal a similar situation: there is a general trend from $\vektor{B}\perp\nabla\Sigma$ to $\vektor{B}\parallel\nabla\Sigma$ (peak at $0^\circ$ to peak at $90^\circ$ in their plots) for increasing column density. However, in their simulations with weaker magnetic field ($10\,\mu\mathrm{G}$) the distribution is very flat at the highest column density with no relevant signal for either orientation.

Using ideal MHD simulations with different turbulent forcing and different magnetic field strengths \citet{KoertgenSoler2020} test the relative orientation in turbulent boxes of non-selfgravitating gas. The turbulent forcing, which is varied from solenoidal to compressive modes, only mildly affects the relative orientation. The variations of the magnetic field strength in their models suggests that only dynamically relevant field strengths can lead to a change in orientation. Compared to our simulations this could mean that the gravitationally driven compression mainly leads to a change in orientation due to the accompanying adiabatic amplification of the field rather than the compressive driving modes. Similar set-ups by \citet{BarretoMotaEtAl2021} cover different sonic and Alfv\'{e}nic Mach numbers. Their results are consistent with our model concerning the orientation as a function of density.

\section{Conclusions}
\label{sec:conclusion}

We perform high resolution simulations of the multi-phase SN-driven interstellar medium in a box of $(0.5\,\mathrm{kpc})^3$. We solve the equations of ideal magnetohydrodynamics including an external potential as well as self-gravity using the MPI parallel code \textsc{Flash}. Radiative cooling is computed using a chemical network that actively follows the abundances of ionized, atomic and molecular hydrogen as well as CO and free electrons. We perform several simulations with increasing resolution from 2 down to $0.25\,\mathrm{pc}$. Our results can be summarized as follows.
\begin{itemize}
    \item We find a clear correlation of the angle between the gradient of the density, $\nabla\rho$, and the magnetic field vector, $\vektor{B}$. From low to high densities, the distribution changes from preferentially perpendicular ($\nabla\rho\perp\vektor{B}$) to preferentially parallel ($\nabla\rho\parallel\vektor{B}$). The transition between the two regimes occurs at a density of approximately $\rho\sim10^{-22}-10^{-21}\,\mathrm{g\,cm}^{-3}$ ($n\sim40-400\,\percc$), see Fig.~\ref{fig:xi-3D}. This transition also coincides with the transition from gas flow perpendicular to the magnetic field vector ($\vektor{v}\perp\vektor{B}$) to a parallel flow ($\vektor{v}\parallel\vektor{B}$) and also with the transition from magnetically supported to gravitationally dominated dense clouds, see Fig.~\ref{fig:grav-ratio}. This transition is indicated by exceeding the critical mass-to-flux ratio as well as a dominating contribution of the gravitational forces compared to the opposing pressure forces (Fig.~\ref{fig:mass-to-flux-ratios}).
    \item This angle measurement is consistent with the hypothesis that the large scale structure (the filaments as a whole) is compressed together with the field lines from the diffuse gas. In high density regions the contracting condensation with a typical size of a few up to 10\,pc form as a result of gravitational attraction in a strongly magnetized environment. This results in gas accretion along the field lines. The dense gas structures build up perpendicular to the field lines, i.e. the density gradients are parallel to the field, see right-hand panel of Fig.~\ref{fig:dens-mag-streamlines-zoom}.
    \item For the projected quantities (column density and projected plane of the sky magnetic field) the correlated orientation can be difficult to measure, i.e. the relative orientation of the gradient of the column density and the vector of the projected magnetic field in the plane of the sky can sensitively depend on the line of sight, the orientation of the cloud and the dynamics in the system under consideration. This includes temporal fluctuations without a clear signal as a function of column density.
We find that the correlation lengths of the magnetic field is comparable or smaller than the typical integration lengths for the dense gas with $\rho\gtrsim10^{-22}-10^{-21}\,\mathrm{g\,cm}^{-3}$ ($n\gtrsim40-400\,\percc$), in which we measure a clear signal in orientation in three dimensions. This indicates that the magnetic field can change its orientation significantly along the line of sight and the measured correlation can sensitively depend on which magnetic field configuration along the line of sight is actually dominating the measurement.
We also analyse the dispersion of the orientation of $B$ along the line of sight and find a broad range of measurements. Along some lines of sight the field orientation is basically unchanged and stable and meaningful quantifications of the relation between field and gas structures can be derived. In other regions the magnetic field changes its orientation so strongly that the resulting projected number does not reflect the possibly universal 3D orientation at all.
In particular for higher resolution simulations, in which the turbulent dynamics can cascade to smaller spatial scales and stir the gas together with the magnetic field, the projected 2D signal might fluctuate or cancel out.
    \item Both the total ratio of gravitational to total pressure forces as well as the orientation between contracting gravitational forces and opposing pressure forces clearly indicate that the dense regions are dominated by gravitational collapse, see Figs.~\ref{fig:gradient-ratios} and \ref{fig:angle-acc-Ptot-PDF}.
\end{itemize}

\section*{Acknowledgements}
We thank the anonymous referee for a very careful reading of the manuscript and valuable comments that helped improving the paper. We thank Thomas Berlok, Daniel Seifried, Thorsten Naab, Stefanie Walch-Gassner, Ralf Klessen, Simon Glover, Richard W\"{u}nsch as well as Martin Sparre and Georg Winner for helpful discussions.
PG acknowledges funding from the European Research Council under ERC-CoG grant CRAGSMAN-646955. The software used in this work was developed in part by the DOE NNSA ASC- and DOE Office of Science ASCR-supported Flash Center for Computational Science at the University of Chicago. Parts of the analysis is carried out using the \textsc{YT} analysis package (\citealt{TurkEtAl2011}, \url{yt-project.org}).

\section*{Data Availability}

The simulation data is publicly available at \href{http://silcc.mpa-garching.mpg.de}{http://silcc.mpa-garching.mpg.de} as DR8. The analysis scripts will be shared upon request.





\bibliographystyle{mnras}
\bibliography{astro.bib,girichidis.bib}



\appendix

\section{Density, magnetic field and mass-to-flux ratio}
\label{sec:app-mass-to-flux-ratio}
The initial density profile is given by
\begin{equation}
    \rho(z) = \rho_0 \exp\left(-\frac{z^2}{2h_\mathrm{gas}^2}\right)
\end{equation}
with a central density of $\rho_0=9\times10^{-24}\,\mathrm{g\,cm}^{-3}$, and a scale height of $h_\mathrm{gas}=30\,\mathrm{pc}$. The column density is thus given by
\begin{align}
    \Sigma &= \int_{z_0}^{z_1} \rho(z) \,\mathrm{d}z\\
    &= \rho_0\,h_\mathrm{gas}\sqrt{\frac{\pi}{2}}\,\left[\mathrm{erf}\left(\frac{z_1}{\sqrt{2}\,h_\mathrm{gas}}\right)-\mathrm{erf}\left(\frac{z_0}{\sqrt{2}\,h_\mathrm{gas}}\right)\right],
\end{align}
integrated from $z_0=-L_z/2=-250\,\mathrm{pc}$ to $z_1=L_z/2=250\,\mathrm{pc}$. This yields a numerical value of $\Sigma=0.00209\,\gperscm$ and a total mass in the simulation box of $M=L_xL_y\Sigma=2.5\times10^6\,\mathrm{M}_\odot$, where $L_x=L_y=L_z=500\,\mathrm{pc}$.
The magnetic field strength scales as the square root of the density with a central field strength of $B_{x,0}=3\,\mu\mathrm{G}$,
\begin{equation}
    B_x(z) = B_{x,0} \exp\left(-\frac{z^2}{4h_\mathrm{gas}^2}\right).
\end{equation}
For the magnetic flux in the $x$ direction we find
\begin{align}
    \Phi &= L_y\,\int_{z_0}^{z_1} B_x(z) \,\mathrm{d}z\\
    &= L_y\,B_0\,h_\mathrm{gas}\sqrt{\pi}\,\left[\mathrm{erf}\left(\frac{z_1}{2\,h_\mathrm{gas}}\right)-\mathrm{erf}\left(\frac{z_0}{2\,h_\mathrm{gas}}\right)\right]\\
    &= 1.519\times10^{36}\,\mathrm{G\,cm}^2
\end{align}
The critical mass-to-flux ratio is
\begin{equation}
\mu_\mathrm{crit} \equiv \left(\frac{M}{\Phi}\right)_\mathrm{crit} = \frac{c_1}{3\pi}\left(\frac{5}{G}\right)^{1/2},
\end{equation}
where $c_1\approx0.53$. For the values in our simulations the global mass-to-flux ratio is $(M/\Phi)/\mu_\mathrm{crit}=6.72$, so it is globally not supported by the magnetic field.

\section{Resolution study for $\xi_\mathrm{3D}$}
\label{sec:app-resolution}
\begin{figure}
    \centering
    \includegraphics[width=8cm]{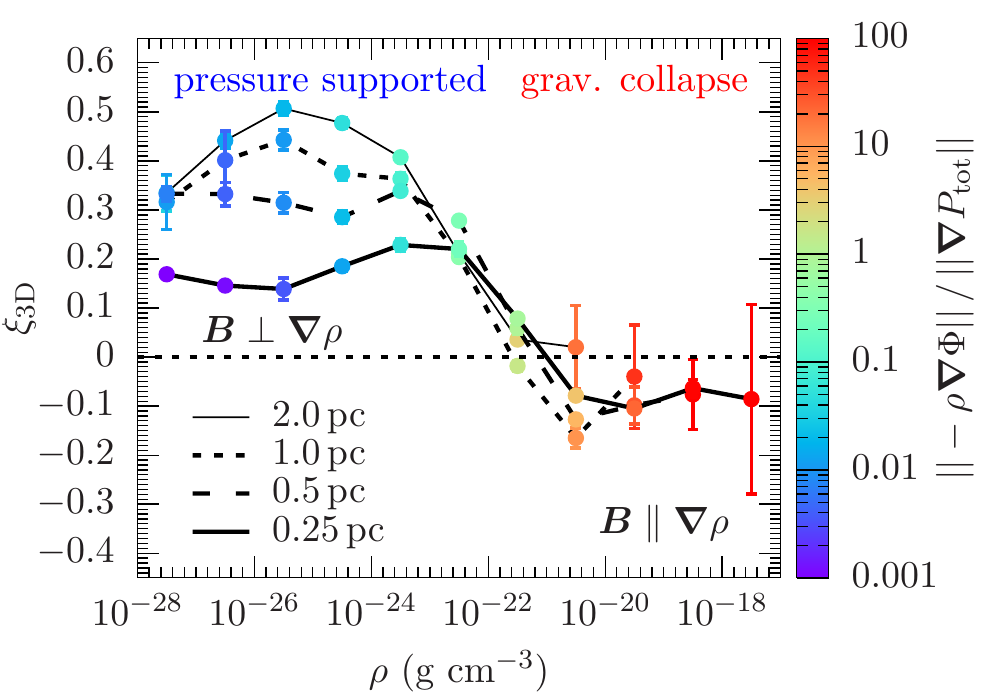}
    \caption{Resolution study of $\xi_\mathrm{3D}$ for resolutions from $2\,\mathrm{pc}$ down to $0.25\,\mathrm{pc}$. At low densities (pressure dominated regime) the values are not converged with resolution. At high densities only the simulations below $1\,\mathrm{pc}$ reliably reach a gravitationally dominated regime with $\xi_\mathrm{3D}<0$.}
    \label{fig:xi-3d-resolution}
\end{figure}

We compute the dependence of $\xi_\mathrm{3D}$ for all resolutions from $2\,\mathrm{pc}$ down to $0.25\,\mathrm{pc}$. Fig.~\ref{fig:xi-3d-resolution} shows the values as a function of density, averaged over the analysis time. For low densities the curves are not converged. None the less, the transition from the pressure supported to the gravitationally dominated regime is very similar for all resolutions.

\section{Magnetic tension}
\label{sec:app-magnetic-tension}

\begin{figure}
    \centering
    \includegraphics[width=8cm]{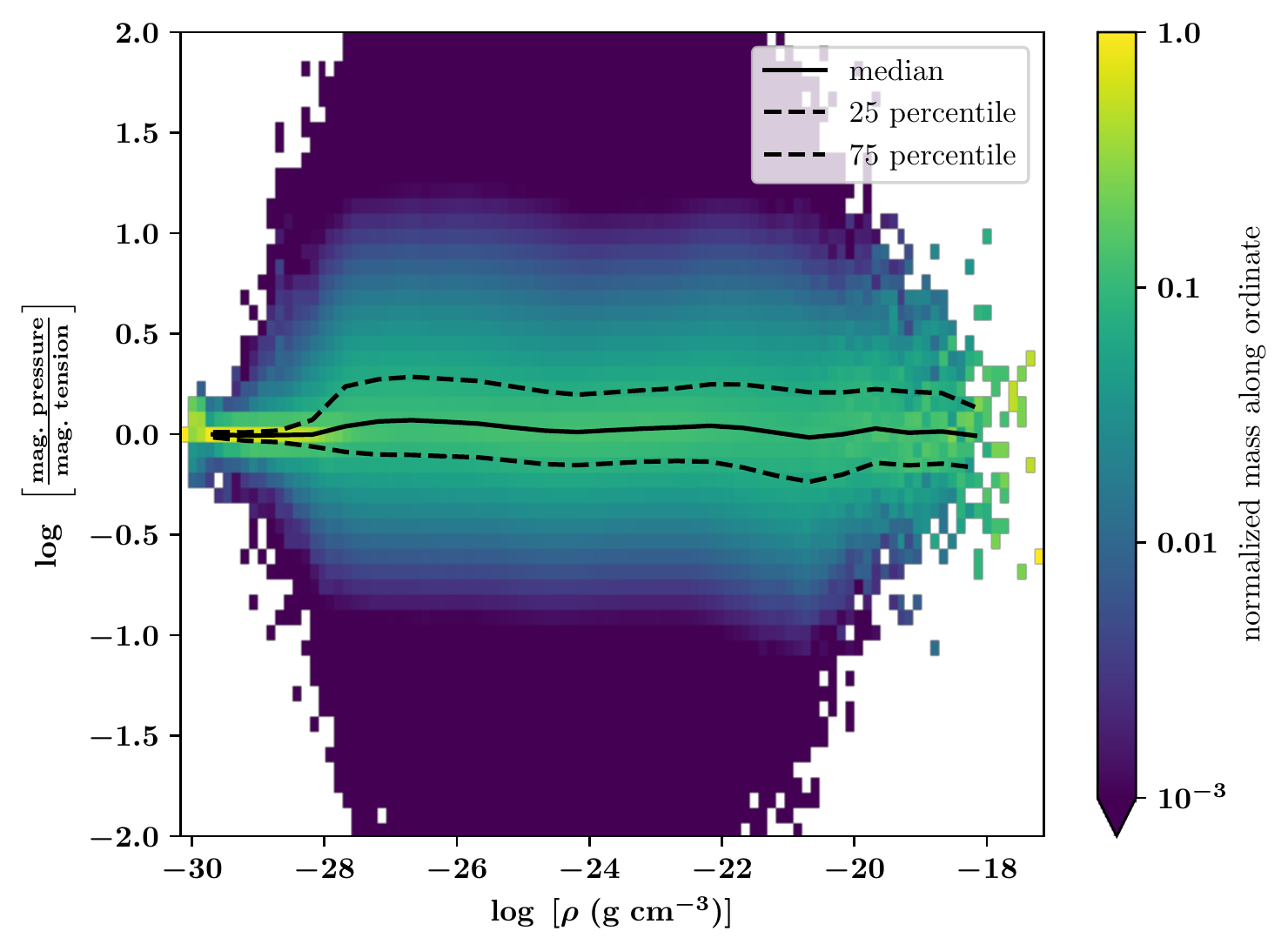}
    \caption{Ratio of magnetic pressure over modulus of the magnetic tension as a function of density in run \texttt{B3-0.25pc} at the end of the simulation time. The distribution is close to unity with the 25 and 75 percentile spanning a factor of $\sim2$.}
    \label{fig:app-magnetic-tension}
\end{figure}

In order to simplify the magnetic analysis we only consider the magnetic pressure and ignore the magnetic tension effects. This however requires the magnetic tension to be at most of the order of the magnetic pressure. Fig.~\ref{fig:app-magnetic-tension} shows the ratio of the magnetic pressure to the modulus of the magnetic tension as a function of density for run \texttt{B3-0.25pc} at the end of the simulation time. Colour coded is the mass normalized along the ordinate for each density bin. The distribution is close to unity for all densities. The 25 and 75 percentile are enclosing a dynamical range of order $2$.

\section{Resolution study for the centroid method}
\label{sec:centroid-resolution}

\begin{figure*}
\begin{minipage}{\textwidth}
\includegraphics[width=\textwidth]{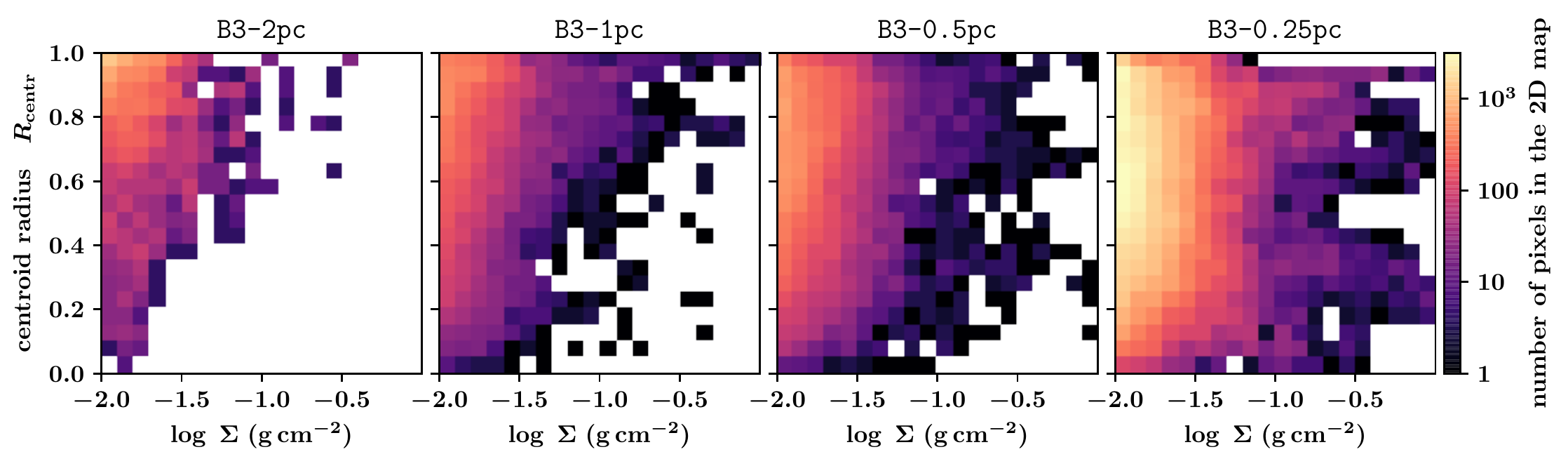}
\caption{Two-dimensional distributions of the centroids as a function of centroid radius and column density for all resolutions. We use all data from the analysis time range.}
\label{fig:centroid-resolution}
\end{minipage}
\end{figure*}

We show two-dimensional distributions of the number of centroid as a function of the centroid radius $R_\mathrm{centr}$ and the column density $\Sigma_\mathrm{los}$ in Fig.~\ref{fig:centroid-resolution}. At lower resolutions there is an overdensity of centroids at larger radii. This is not surprising since there are fewer cells along the line of sight and strong changes in the field configuration along only a few cells is less likely. For simulations \texttt{B3-0.5pc} and \texttt{B3-0.25pc} the distributions cover the entire range in centroid radius from zero (very broad distribution in angles) to unity (no significant change of field configuration along the line of sight). As a result the low resolution simulations might show a more stable correlation between the orientation of the magnetic field with respect to the column density structure.


\bsp	
\label{lastpage}
\end{document}